\documentclass{aa}  
\usepackage{comment}
\usepackage{siunitx}
\usepackage{subcaption}
\usepackage{nicefrac}
\usepackage{graphicx}
\usepackage{color}
\usepackage{hyperref}
\usepackage{csvsimple}
\usepackage{ulem}
\usepackage[switch]{lineno}

\usepackage{booktabs,wrapfig}
\hypersetup{colorlinks,linkcolor={blue},citecolor={blue}} 
\usepackage[dvipsnames,table,xcdraw]{xcolor}

\usepackage{textcomp,gensymb}
\usepackage{caption} 
\usepackage{tabularx}
\usepackage[bottom]{footmisc}
\usepackage{mathtools}
\usepackage[varg]{txfonts}
\begin{document} 

    \title{A Galactic intermediate-mass stripped star with a Wolf--Rayet-like wind}

   \author{Johanna M\"uller-Horn\inst{\ref{inst:mpia}, \ref{inst:uni-hd}}\thanks{Corresponding author \email{mueller-horn@mpia.de}} \and
          Kareem El-Badry \inst{\ref{inst:caltech},\ref{inst:mpia}} \and 
          Andreas A. C. Sander \inst{\ref{inst:ari},\ref{inst:iwr}} \and 
          Hans-Walter Rix \inst{\ref{inst:mpia}} \and
          Lisa Blomberg \inst{\ref{inst:caltech}} \and 
          J. J. Hermes \inst{\ref{inst:boston}} \and
          Pranav Nagarajan \inst{\ref{inst:caltech}} \and 
          Sahar Shahaf \inst{\ref{inst:mpia}} \and 
          Harim Jin \inst{\ref{inst:mpa}} \and 
          Dominick M. Rowan \inst{\ref{inst:ucb}} \and 
          Debasish Dutta \inst{\ref{inst:ista}} \and 
          Jos\'e G. Fern\'andez-Trincado \inst{\ref{inst:Santiago}} \and 
          Ylva G\"otberg \inst{\ref{inst:ista}} \and 
          Ilya Ilyin \inst{\ref{inst:aip}} \and
          Tom Maccarone \inst{\ref{inst:utt}} \and 
          Jos\'e Eduardo M\'endez Delgado \inst{\ref{inst:unam}}\and
          Guy S. Stringfellow \inst{\ref{inst:colorado}} \and
          Andrew Tkachenko \inst{\ref{inst:KUL}} \and
          Jaime I. Villase\~nor \inst{\ref{inst:mpia}} \and  
          Eleonora Zari \inst{\ref{inst:firenze}}
          }

   \institute{
    {Max-Planck-Institut für Astronomie, Königstuhl 17, 69117 Heidelberg, Germany \label{inst:mpia}} 
    \and {Fakultät für Physik und Astronomie, Universität Heidelberg, Im Neuenheimer Feld 226, 69120 Heidelberg, Germany \label{inst:uni-hd}}
    \and {Department of Astronomy, California Institute of Technology, Pasadena, CA 91125, USA \label{inst:caltech}}
    \and {Zentrum für Astronomie der Universität Heidelberg, Astronomisches Rechen-Institut, Mönchhofstr. 12-14, 69120 Heidelberg, Germany \label{inst:ari}}
    \and {Interdisziplin{\"a}res Zentrum f{\"u}r Wissenschaftliches Rechnen, Universit{\"a}t Heidelberg, Im Neuenheimer Feld 225, 69120 Heidelberg, Germany\label{inst:iwr}}
    \and {Department of Astronomy \& Institute for Astrophysical Research, Boston University, 725 Commonwealth Ave., Boston, MA 02215, USA \label{inst:boston}}
    \and {Max-Planck-Institut für Astrophysik, Karl-Schwarzschild-Strasse 1, 85748 Garching, Germany \label{inst:mpa}}
    \and {Department of Astronomy, University of California, Berkeley, CA 94720, USA \label{inst:ucb}}
    \and {Institute of Science and Technology Austria (ISTA), Am Campus 1, 3400 Klosterneuburg, Austria\label{inst:ista}}
    \and {Centro de investigaci\'on en Astronom\'ia, Facultad de Ingenier\'ia, Ciencia y Tecnolog\'ia, Universidad Bernardo O’Higgins, Av. Viel 1497, Santiago, 8370993, Chile \label{inst:Santiago}}
    \and {Leibniz-Institut für Astrophysik Potsdam (AIP), An der Sternwarte 16, 14482 Potsdam, Germany \label{inst:aip}}
    \and {Department of Physics \& Astronomy, Texas Tech University, Box 41051, Lubbock, TX, 79409-1051, USA \label{inst:utt}}
    \and {Instituto de Astronom\'ia, Universidad Nacional Aut\'onoma de M\'exico, A.P. 70-264, 04510, Mexico, D.F., M\'exico \label{inst:unam}}
    \and University of Colorado Boulder, Boulder, CO 80309  USA \label{inst:colorado}
    \and {Institute of Astronomy, KU Leuven, Celestijnenlaan 200D, 3001 Leuven, Belgium \label{inst:KUL}}
    \and {INAF–OAA, Osservatorio Astrofisico di Arcetri, largo E. Fermi 5, 50127, Firenze, Italy \label{inst:firenze}}
    }

   \date{Received 04 August 2026}

 
  \abstract
    {Binary interaction in massive stars is expected to produce a large population of intermediate-mass ($2$–$8\,\mathrm{M}_\odot$) envelope-stripped stars, yet such objects have remained elusive in the Milky Way. We report the identification of an unambiguous Galactic example in a short-period ($P = 5.94$ d), double-lined spectroscopic binary, discovered in the SDSS-V Milky Way Mapper survey. The system consists of a rapidly rotating O-type star and a hotter, lower-mass companion, which shows \ion{He}{ii} and \ion{N}{iv} emission lines with large radial velocity variations, revealing its binary nature.
    
    Combined orbital constraints and joint spectroscopic and photometric modelling show that the companion is a hot ($T_\ast \approx 60\,\mathrm{kK}$), helium-rich star with a mass of $3.2$--$5.8\,\mathrm{M}_\odot$, placing it squarely in the intermediate-mass regime and below values typically inferred for classical Wolf--Rayet (WR) stars.
    Its spectral properties are inconsistent with accretion-powered emission and instead indicate a stripped star formed through binary mass transfer. The system’s short period, negligible eccentricity, and rapidly rotating O-star point to a post-interaction configuration following efficient mass transfer and spin-up of the accretor.
    
    Comparison with binary evolution models suggests that the stripped star is observed in a brief inflated phase following mass transfer, which increases its optical flux contribution and facilitates its detection. The inferred mass-loss rate $\log \dot{M} / (\mathrm{M}_\odot \mathrm{yr}^{-1})= -6.3 \pm 0.1$ is in line with mass-loss rates observed for classical WR stars in the Milky Way and exceeds those measured for intermediate-mass stripped stars in the Magellanic Clouds, with the caveat that our target selection is biased towards systems with stronger emission features.
    
    As an unambiguous and well-characterised intermediate-mass stripped star, this system provides a key benchmark for models of binary evolution at solar metallicity, stripped-envelope supernova progenitors, and the formation of compact-object binaries.} 
   \keywords{Binaries: spectroscopic, Stars: individual: {\textit{Gaia} DR3 524993029624315904}, Stars: early-type, Stars: Wolf-Rayet}

   \maketitle

\section{Introduction}
Binary interaction is fundamental to the evolution of massive stars. Observational studies indicate that up to $\sim70\%$ of massive stars experience binary interaction during their lifetimes, with a substantial fraction ($\sim30\%$) undergoing envelope stripping via Roche-lobe overflow \citep[e.g.][]{Sana+2012, deMink+2014}. This process removes the hydrogen-rich envelope of the donor star, leaving behind a hot, compact, helium-rich object. Such stripped stars form across a wide range of masses, from low-mass hot subdwarf B- and O-type stars \citep[$\lesssim1.5\,\mathrm{M}_\odot$; e.g.][]{Heber2016,Han+2002,Han+2003} to massive classical Wolf-Rayet (WR) stars \citep[$\gtrsim8\,\mathrm{M}_\odot$; e.g.][]{Shenar+2016, Shenar+2018, SanderVink2020}, with thousands of hot subdwarfs and several hundred confirmed WR stars observed in the Milky Way \citep{Culpan+2023,Rosslowe_Crowther2015}\footnote{ \href{http://pacrowther.staff.shef.ac.uk/WRcat/index.php}{http://pacrowther.staff.shef.ac.uk/WRcat/index.php}}.

Stripped stars are important across several domains of astrophysics. Those with helium core masses above $\gtrsim2.6\,
\mathrm{M}_\odot$ (roughly corresponding to initial masses $\gtrsim 10\,\mathrm{M}_\odot$) are expected progenitors of hydrogen-poor core-collapse supernovae \citep[e.g.][]{Eldrige+2013,Tauris+2015,Laplace+2021, Chanlaridis+2022}. Stripped stars hence represent a crucial intermediate stage in the formation of compact-object binaries, including double neutron stars and black hole systems, which directly links them to the population of gravitational-wave sources \citep[e.g.][]{DeDonder_Vanbeveren1998,Tauris+2017,Vigna-Gomez+2018, Mandel_Broekgaarden2022}. Their high effective temperatures and bright spectra in the ultraviolet (UV) also make them efficient emitters of ionising radiation, with potentially significant contributions to the ionising photon budget of galaxies \citep{Stanway+2016,Goetberg+2019,Doughty_Finlator2021}. More broadly, the observed properties of stripped stars and their companions encode key information about poorly constrained phases of binary evolution, including the stability and efficiency of mass transfer and the physics of common-envelope evolution \citep{Lechien+2025, Picco+2026, Seeburger+2026, Nagarajan+2026, Sen+2026}. This makes them essential benchmarks for calibrating binary population synthesis models.

Stars with initial masses between approximately $8$ and $25\,
\mathrm{M}_\odot$ are predicted to produce a population of intermediate-mass stripped stars with masses in the range of $\sim$2--$8\,
\mathrm{M}_\odot$ \citep[e.g.][]{Goetberg+2018,Hovis-Afflerbach+2025}. These objects occupy the evolutionary gap between low-mass hot subdwarfs and classical WR stars and are predicted to be hot ($T_\mathrm{eff} \approx 20$--$100$\,kK), compact ($R_* \lesssim 1\,\mathrm{R}_\odot$), and relatively long-lived \citep[with the core helium burning stage lasting for about $10\%$ of the total life time of the star;][]{Goetberg+2018,DuttaKlencki2024}. Binary population synthesis studies suggest the Milky Way may host $\sim3\times10^3$--$6\times10^4$ such systems \citep{Yungelson+2024,Hovis-Afflerbach+2025,Wang+2026}, though these estimates remain uncertain due to the complex interplay of mass transfer, stellar winds, binary dynamics, and stellar evolution in general. Although much of the theoretical effort has been focused on modelling the physics of donor stars, the accreting companions are also expected to be significantly affected by mass transfer; they may be spun up to near-critical rotation and chemically altered by accretion of helium-rich and CNO-processed material \citep{Packet1981, deMink+2013, Renzo+2023, Richards+2025, Jin+2026}.

Despite their predicted abundance, intermediate-mass stripped stars have proven extremely difficult to detect. Their signatures are typically diluted or obscured in the optical by luminous OB-type companions. This observational challenge has motivated alternative detection strategies, particularly searches for UV excess \citep{Goetberg+2018}. Only recently have convincing detections been reported, primarily in the Large and Small Magellanic Clouds (LMC / SMC), revealing objects with $T_\mathrm{eff} \sim 50$--$90\,$kK and luminosities intermediate between hot subdwarfs and WR stars \citep{Drout+2023,Goetberg+2023}, as well as several examples of cooler, partially stripped stars in Be star binaries \citep{Ramachandran+2023,Ramachandran+2024, Villasenor+2023}. Large-scale UV photometric searches have since identified numerous additional candidates \citep{Ludwig+2025,Blomberg+2026}, suggesting a substantial but previously hidden population. 

However, in the Milky Way, clear detections remain scarce. Several relatively massive hot subdwarfs ($\gtrsim1\,
\mathrm{M}_\odot$) have been identified in binaries with rapidly rotating Be stars via UV spectroscopy \citep{Gies+1998, Wang+2021, Mueller-Horn+2026}, and a number of systems initially proposed as OB+black hole binaries have been reinterpreted as containing (often inflated) stripped stars \citep{El-Badry_Quataert2020,Shenar+2020,Bodensteiner+2020,Naze_Rauw2025,Mueller-Horn+2025}. Notably, the long-standing Galactic intermediate-mass stripped star candidate HD~45166 was recently reclassified as a magnetised quasi-WR star formed by the merger of two lower-mass helium stars \citep{Shenar+2023}.

In this work, we report the unambiguous identification of an intermediate-mass stripped star in the Milky Way, discovered with low-resolution optical spectroscopy. We present follow-up optical spectroscopic observations and a detailed analysis of the system's stellar and binary parameters through combined spectroscopic and photometric modelling. This object provides a rare Galactic benchmark for intermediate-mass stripped stars at near-solar metallicity, offering direct observational constraints on mass-loss rates, radii, and ionising fluxes, which are critical inputs for models of binary evolution, supernova progenitors, and compact-object formation.

\section{Target selection and observations}
\label{sec:observations}

The target was identified as part of the Milky Way Mapper (MWM) project within the fifth generation of the Sloan Digital Sky Survey \citep[SDSS-V;][]{Kollmeier+2017,Kollmeier+2026,SDSS_DR19}. The MWM OB star core programme (\texttt{mwm\_ob}) is collecting low-resolution optical spectroscopy for a large sample ($\sim 5\times10^5$) of OBA-type stars, with the aim of mapping the Galactic population of young, hot, and massive stars. The target selection in the \texttt{mwm\_ob} programme is based on simple photometric and astrometric criteria, combining data from \textit{Gaia} (E)DR3 and 2MASS \citep{2MASS}. Effectively, the survey targets luminous ($M_K < -0.6$\,mag, where $M_K$ is the absolute magnitude in the 2MASS $K_s$ band) and blue sources selected via optical and near-infrared colour--magnitude cuts \citep{Zari+2021, Zari+2025}.
The selection is designed to include stars earlier than approximately B3\,V, prioritising completeness over purity.

Using the SDSS-V \texttt{mwm\_ob} observations, we are compiling a homogeneous catalogue of Galactic emission-line stars (Müller-Horn et al., in prep.). The object studied here was identified in this search as one of the few sources exhibiting strong \ion{He}{ii\,$\lambda4686$} emission. Visual inspection of the spectrum revealed additional peculiar features that motivated a detailed analysis and dedicated follow-up observations.
We obtained optical spectroscopy of the target from multiple facilities: 

As part of the SDSS-V, three spectra were taken with the Baryon Oscillation Spectroscopic Survey spectrograph \citep[BOSS;][]{Smee+2013} mounted on the 2.5\,m telescope at Apache Point Observatory (APO), New Mexico \citep{Bowen_Vaughan1973, Gunn+2006}. The BOSS spectra cover the wavelength range $3622$--$10354$\,\AA\ with a spectral resolving power of $R \simeq 1800$. Data were processed using the standard BOSS reduction pipeline (version \texttt{v6\_2\_1}; Morrison et al., in prep.; Johnson et al., in prep).

In addition, we obtained two high-resolution spectra with the High Resolution Echelle Spectrometer \citep[HIRES;][]{Vogt+1994} on the 10\,m Keck~I telescope at the W.~M.~Keck Observatory, Maunakea. The spectra were observed in the HIRESb configuration, have a resolving power of $R \simeq 55\,000$, and cover the wavelength range $\sim 3100$--$5900$\,\AA. Data were reduced using the \href{https://sites.astro.caltech.edu/~tb/makee/}{\texttt{MAKEE}} software.

We further obtained four long-slit spectra with the GMOS-N instrument on the 8.1\,m Gemini North telescope, Maunakea (programme \texttt{GN-2025B-Q-420}, PI P.~Nagarajan). The observations were carried out using the R831 grating with a 0.5\,arcsec slit and a central wavelength of 650\,nm, resulting in a wavelength coverage of $5400$--$7500$\,\AA\ and a resolving power of $R \simeq 4000$. These data were reduced using the \texttt{DRAGONS} package (version \texttt{v4.1.0}).

We also obtained a single-epoch spectrum using the high-resolution PEPSI spectrograph \citep{Strassmeier+2015} at the 8.4\,m Large Binocular Telescope (LBT) at the Mount Graham International Observatory, Arizona (PI D.~Rowan). Observations were carried out in the CD\,2, 3, 4, and 5 channels, covering the wavelength range $4200$--$7200$\,\AA\ at a resolving power of $R \simeq 50\,000$. The 2D echelle spectrum was processed following the procedure outlined in \citet{Strassmeier+2018}. Due to the high extinction towards the source and the resulting low signal-to-noise ratio (S/N) in the blue arm, we restricted the subsequent analysis to wavelengths $\gtrsim 5000$\,\AA.

All 10 spectra were corrected for barycentric motion and were continuum-normalised. We continuum-normalised each spectrum using an iterative sigma-clipping and spline-fitting procedure. After selecting pixels with valid flux measurements and non-zero inverse variance, we constructed an initial continuum estimate using a broad running high-percentile filter. This estimate is then refined by iteratively clipping deviant pixels and fitting a cubic spline to the remaining continuum points. 
A summary of the observations, including dates and typical S/N, is provided in Table~\ref{tab:obs_overview}.

\section{First spectral clues for a post-interaction binary}
\label{sec:first_look}

\begin{figure}[t!]
    \centering
    \includegraphics[width=\columnwidth]{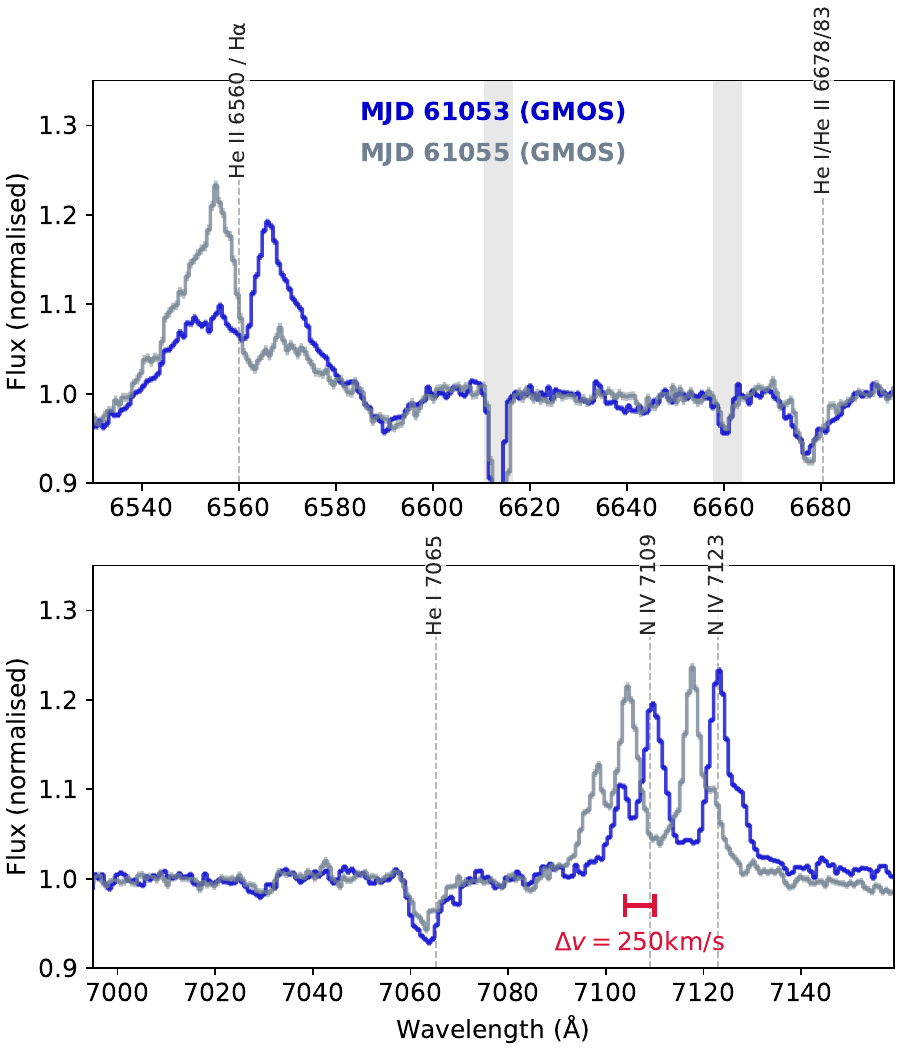}
    \caption{Short-term spectroscopic variability. Selected wavelength regions from two of the GMOS spectra taken two days apart, around the Balmer H$\alpha$ line (top) and the \ion{N}{iv\,$\lambda\lambda\,7103$--$7109$--$7123$} lines (bottom). Line labels indicate the rest-frame wavelengths of diagnostic lines. Interstellar absorption lines are marked with grey shaded bands. A large Doppler shift is seen in the \ion{N}{iv} emission lines, whereas the \ion{He}{I} lines remain close to stationary. The H$\alpha$ line shows strong variability in the line profile; line profile changes are visible to a lesser extent also in the \ion{He}{i}/\ion{He}{ii} lines near 6680\,\AA. } 
    \label{fig:spectral_variability}
\end{figure} 

The target system, \textit{Gaia} DR3 524993029624315904 (SDSS ID 55440347) is a faint ($G = 14.01$\,mag) star located in the outer Galactic disc ($\alpha, \delta =$ 01:11:58.49, +64:49:45.35). As we will show in Sect.~\ref{sec:spectral_fit_results}, its spectrum exhibits the characteristic emission-line signatures of a WR star. Since the system was not included in previous editions of the Galactic WR catalogue, we assign it the designation WR\,2-1 following the naming convention described in \citet{Rosslowe_Crowther2015}.\footnote{The identifier WR\,2-1 denotes the first newly identified Galactic WR star with right ascension between known WR stars WR\,2 and WR\,3.}

Fig.~\ref{fig:spectral_variability} shows two GMOS spectra obtained two days apart, highlighting key spectral features that motivated and guided the subsequent analysis. The spectra display the characteristic absorption features of OB-type stars, including broad Balmer and Paschen lines and numerous \ion{He}{i} and \ion{He}{ii} absorption lines. The \ion{He}{i} lines appear to be nearly stationary and do not show strong variations in radial velocity (RV).
In addition, the spectra exhibit prominent emission features that are not expected for typical OB dwarfs. These include strong emission in H$\alpha$, \ion{He}{ii\,$\lambda\,4686$} and \ion{He}{ii\,$\lambda\,10125$}, as well as high-ionisation nitrogen lines such as \ion{N}{iv\,$\lambda\,4058$} and \ion{N}{iv\,$\lambda\lambda\,7103$,$7109$,$7123$}. Weak P-Cygni profiles are also visible in \ion{N}{v\,$\lambda\lambda\,4604/20$}. In contrast to the absorption lines, these emission features exhibit large RV shifts, exceeding $\sim 250$\,km\,s$^{-1}$ on timescales of a few days (Fig.~\ref{fig:spectral_variability}).

These combined signatures are difficult to reconcile with the spectrum of a single star because the features point to distinct temperature regimes: The \ion{He}{i}/\ion{He}{ii} absorption spectrum is consistent with a late O-type star ($T_{\ast} \sim 30$--$40$\,kK), whereas the presence of strong \ion{N}{iv} and \ion{N}{v} lines requires substantially higher temperatures ($\gtrsim 50$--$60$\,kK). The discrepant RV behaviour provides further evidence for multiple components in the spectra. 

Together, these properties indicate a composite binary spectrum arising from two stars of different temperatures. A possible interpretation is a post-interaction configuration comprising a massive OB-type star and a hotter, but less massive stripped star companion, given the larger RV variation of the hotter star.

\section{Spectroscopic binary orbit}

We quantified the binary nature of the system by measuring RVs for both components and constraining the orbital solution. Here, we refer to the OB star dominating the \ion{He}{i} absorption as the primary, and the emission-line object as the secondary or companion.

\subsection{Radial velocities}
\label{sec:radial_velocities}

We determined RVs of the two components via cross-correlation with synthetic template spectra. We used model spectra computed with the Potsdam Wolf–Rayet (PoWR) code \citep{Graefener+2002,Hamann_Graefener2003,Sander+2015} as the cross-correlation templates for both stars. PoWR provides state-of-the-art non-local thermodynamic equilibrium (non-LTE) models for hot stars with winds \citep[e.g.,][]{Sander+2024}. 

As a template for the primary, we adopted a model from the OB-star Milky Way grid \citep{Hainich+2019} with $T_\ast = 36\,\mathrm{kK}$, $\log g = 4.2$, $\varv_\mathrm{rot} \sin i = 200\,$km/s, and solar metallicity. For the secondary, we used a PoWR model from the WR grid as the cross-correlation template, with $T_\ast = 56\,\mathrm{kK}$ and $R_\mathrm{t} = 40\,\mathrm{R}_\odot$. The choice of templates was informed by our first assessment from Sect.~\ref{sec:first_look} and the fit results later in the article, see Sect.~\ref{sec:spectral_fit_results}.

The cross-correlation was performed in multiple wavelength regions, depending on the available wavelength coverage for each spectrum. For the primary, we used the \ion{He}{i} absorption lines at $\lambda\lambda\,4471,\,5876,$ and $7065$, which are expected to have only a minor contribution from the hot companion (as opposed to \ion{He}{ii} lines, for example). For the companion star, we used wavelength regions centred on the \ion{N}{iv\,$\lambda\lambda$\,4058,\,7110} emission lines, which provide the clearest tracers of its orbital motion.

The resulting RVs for both components are listed in Table~\ref{tab:obs_overview}. RVs and uncertainties were estimated following \citet{Zucker2003}, with uncertainty estimates based on the curvature of the peak of the cross-correlation function. As a consistency check, we repeated the RV analysis using the final best-fit model spectra together with two-dimensional template cross-correlation. This independent analysis yielded RVs consistent with those from the one-dimensional approach (Appendix~\ref{appendix:todcor}).

\subsection{Orbital analysis}
\label{sec:binary_orbit}

\begin{table}[t]
\centering
\caption{Orbital parameters and uncertainties.}
\renewcommand{\arraystretch}{1.6}
\begin{tabular}{l c c}
\hline
\hline
Orbital period & $P$ [d] &  $5.944_{-0.007}^{+0.009}$\\ 
Eccentricity & $e$ & $0$  \textit{(fixed)}\\ 
Barycentric velocity & $v_z$ [km s$^{-1}$] & $-79.9\pm5.1$ \\
Time of conjunction & $T_\mathrm{conj}$ [MJD] & $61056.9\pm0.1$ \\
Dynamic mass ratio & $q_\mathrm{dyn}=\frac{K_\mathrm{2}}{K_1}=\frac{M_\mathrm{1}}{M_2}$ & $5.8_{-1.5}^{+2.8}$ \\
RV semi-amplitude &  & \\ 
-- Primary  & $K_{\mathrm{1}}$ [km s$^{-1}$] & $25.4_{-8.1}^{+7.7}$\\ 
-- Companion  & $K_{\mathrm{2}}$ [km s$^{-1}$] & $149.2_{-8.0}^{+7.5}$\\ 
Dynamic mass&  & \\
-- Primary  & $M_{\mathrm{1}} \sin^3 i$ [M$_\odot$] & $2.81_{-0.45}^{+0.47}$ \\ 
-- Companion  & $M_{\mathrm{2}} \sin^3 i$ [M$_\odot$] & $0.48_{-0.18}^{+0.21}$\\ 
RV jitter & & \\
-- Primary & $s_1$ [km s$^{-1}$] & $17_{-5}^{+6}$\\ 
-- Companion & $s_2$ [km s$^{-1}$] & $9_{-7}^{+9}$\\
\hline
\end{tabular}
\label{tab:orbit_parameters}
\setlength{\extrarowheight}{0pt}
\end{table}

\begin{figure}[ht!]
    \centering
    \includegraphics[width=\columnwidth]{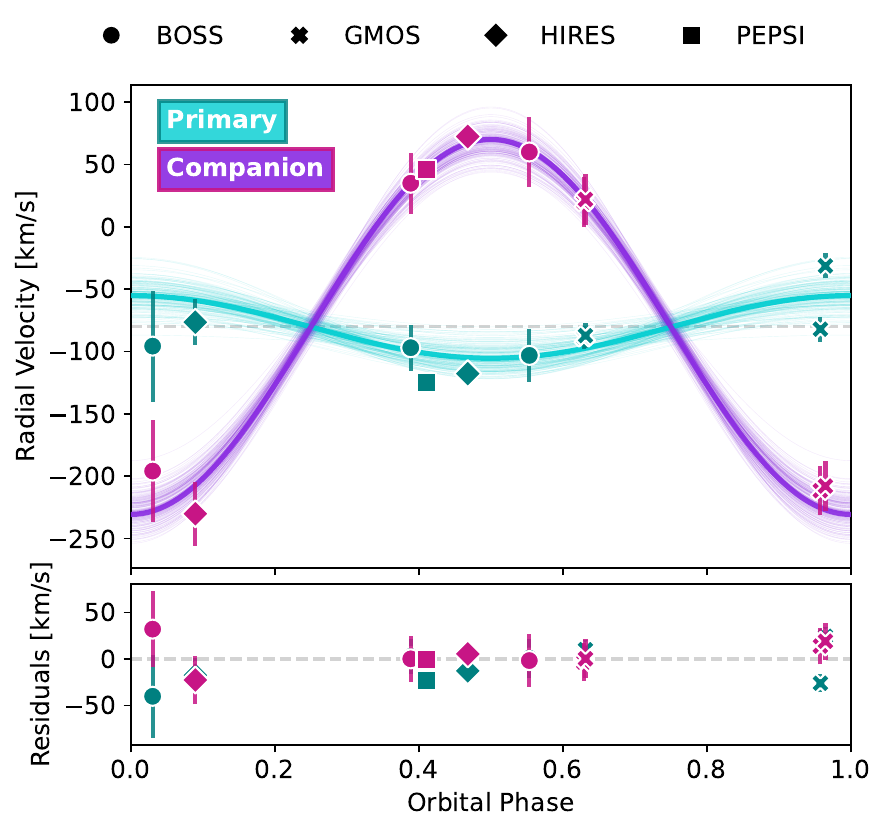}
    \caption{Phase-folded RV curves of both components of the WR\,2-1 binary. Measurements of the primary (teal) and companion (pink) are shown with symbols indicating the different instruments. The best-fit orbital solution is overplotted for the primary (cyan) and companion (purple). Thin lines show random posterior samples to illustrate parameter uncertainties. The horizontal dashed line marks the systemic velocity.
    } 
    \label{fig:orbit_fit}
\end{figure} 

We inferred the orbital parameters by jointly fitting the RV curves of both components using a nested sampling approach. We used the \texttt{MLFriends} algorithm \citep{Buchner2016,Buchner2019}, as implemented in \texttt{UltraNest} \citep{Buchner2021}, to sample the posterior distributions.

A double-lined spectroscopic binary is described by seven parameters: the RV semi-amplitudes ($K_1$, $K_2$), orbital period ($P$), eccentricity ($e$), argument of periastron ($\omega$), systemic velocity ($v_z$), and a phase parameter ($\tau$). Following \citet{Blunt+2020}, we define $\tau$ relative to a reference time $T_\mathrm{ref} = \mathrm{MJD}\,51544$ as
$\tau = \left(\frac{T_\mathrm{ref} - T_\mathrm{peri}}{P}\right) \bmod 1 \,.$ To account for additional scatter, for example due to template mismatch, we included RV jitter terms ($s_1$, $s_2$), added in quadrature to the formal uncertainties.
We adopted uniform priors for all parameters: $P \in [1, 100]$\,d; $K_1, K_2 \in [0, 500]$\,km\,s$^{-1}$; $\tau, e \in [0, 1)$; $\omega \in [0, 2\pi]$; $v_z \in [-200, 200]$\,km\,s$^{-1}$; and $s_1, s_2 \in [0, 30]$\,km\,s$^{-1}$. An initial fit allowing for eccentricity yielded values consistent with $e = 0$ within $2\sigma$. We therefore adopted a circular orbit ($e=0$); in this case $\omega$ is not defined, and instead we report a time of conjunction $T_\mathrm{conj}$.

The best-fit solution is shown in Fig.~\ref{fig:orbit_fit}, and the inferred parameters are listed in Table~\ref{tab:orbit_parameters}. We find an orbital period of $5.944_{-0.007}^{+0.009}$\,d, consistent with the short-term RV variability identified in Sect.~\ref{sec:first_look}. The RV semi-amplitudes are $K_1 = 25.4_{-8.1}^{+7.7}$\,km\,s$^{-1}$ and $K_2 = 149.2_{-8.0}^{+7.5}$\,km\,s$^{-1}$ for the primary and companion, respectively, corresponding to a dynamical mass ratio $q_\mathrm{dyn} = K_2 / K_1 = M_1 / M_2 = 5.8_{-1.5}^{+2.8}$.
These values imply minimum masses of $M_1 \sin^3 i = 2.81_{-0.45}^{+0.47}\,M_\odot$ and $M_2 \sin^3 i = 0.48_{-0.18}^{+0.21}\,M_\odot$. The low value obtained for the primary is inconsistent with expectations for an OB-type star, suggesting that the system is observed at a relatively low inclination angle ($i$).

\section{Composite spectroscopic modelling and stellar parameters}
\label{sec:spectral_fit}

To characterise the binary system and test the stripped-star interpretation of the companion, we require robust constraints on the stellar parameters of both components. We therefore performed a joint spectroscopic and photometric analysis, fitting the normalised optical spectra together with the flux-calibrated spectral energy distribution (SED).

A composite modelling approach is required because both stars contribute significantly to the observed spectrum. This is evident from the phase-dependent variability of key diagnostic lines such as \ion{He}{ii} and H$\alpha$, which indicates that the line profiles are blends of contributions from both components. Even diagnostics that are largely dominated by one star, such as \ion{He}{i} absorption, are affected by the other star because its continuum flux contribution will dilute the lines.

In principle, the component spectra could be separated using spectral disentangling techniques \citep[e.g.][]{Simon_Sturm1994, Hadrava+1995, Ilijic+2004, Sablowski+2019, Shenar+2020,Seeburger+2024}. However, this requires multi-epoch observations with homogeneous wavelength coverage, which are not available for the present dataset. We therefore adopt a composite approach, fitting the observations with a combined model of two luminous stars.

\subsection{Observational constraints}

We compiled archival photometry using the VO Sed Analyzer \citep[VOSA;][]{Bayo+2008}, including optical \textit{Gaia} bands \citep[$G$, $G_{BP}$, $G_{RP}$;][]{GaiaDR3}, near-infrared 2MASS data \citep[$J$, $H$, $K_s$;][]{Cutri+2003}, and mid-infrared measurements from WISE \citep[W1--W4;][]{Wright+2010}. The photometric fluxes and uncertainties are listed in Table~\ref{tab:phot_overview}.

For the spectroscopic constraints, we combined the highest S/N and resolution data to achieve broad wavelength coverage. Specifically, we used the HIRES spectrum at MJD~61058 ($\lambda < 5900$\,\AA), the PEPSI spectrum ($5900 \leq \lambda \leq 7200$\,\AA), and the BOSS spectrum at MJD~60636 ($\lambda > 7200$\,\AA). We excluded wavelength windows with gaps in the detectors or strong telluric absorption, restricting the fitted range to selected wavelength windows with diagnostic lines (including hydrogen Balmer, \ion{He}{i / ii}, \ion{N}{iii / iv / v}, and \ion{C}{iv} lines). The HIRES and PEPSI spectra were rebinned (by factors of 15 and 5) to increase S/N values, reducing the effective resolution to $\sim 17\,000$ and $20\,000$, respectively. 

\subsection{Stellar atmosphere models and synthetic photometry}

We modelled both components using non-LTE stellar atmosphere models computed with the PoWR code.
For the primary, we used models from the OB-star Milky Way grid \citep{Hainich+2019} with solar metallicity and varying $T_\ast$ and $\log g$. Based on the observed Balmer line profiles and the simultaneous presence of \ion{He}{i} and \ion{He}{ii}, we restricted the explored parameter space to $30 \leq T_\ast/\mathrm{kK} \leq 45$ and $3.0 \leq \log g \leq 4.5$.

The companion was modelled using PoWR atmospheres tailored to reproduce the observed emission-line spectrum. We obtained initial constraints on the companion parameters by comparing the spectra with the public WN model grids at solar metallicity \citep{Hamann+2004,Todt+2015}. The nitrogen ionisation balance, with strong observed \ion{N}{iv} and weak \ion{N}{iii} and \ion{N}{v} lines, indicates temperatures of $T_\ast \sim 50$--$70$\,kK for the companion. The relatively narrow emission lines suggest comparatively low mass-loss rates ($\log \dot{M} \lesssim -6$). The simultaneous presence and relative strengths of the Balmer and \ion{He}{ii} emission lines require a substantial surface hydrogen fraction. Comparisons with the public PoWR grids favour models with hydrogen mass fractions of approximately $X\approx0.5$ (Appendix~\ref{appendix:hydrogen_frac}).

Based on these preliminary findings, we calculated a series of hydrodynamically-consistent atmosphere models using the PoWR$^\textsc{hd}$ branch \citep{Sander+2017,Sander+2023}. Due to the rather moderate emission lines and the temperature range inferred from the initial grid comparison, the wind of the WR component is not strong enough to be compatible with an immediate launching from the hot iron opacity bump \citep[cf.][]{Sander+2020,Lefever+2025,Lefever+2026}. We therefore use a ``shallow'' modelling approach in this work \citep[cf.][]{Josiek+2025,Lefever+2025} with an inner boundary of $\tau_\text{Ross,cont} = 5$ and add a turbulent pressure into the hydrodynamic equation described by a constant velocity of $\varv_\text{turb} = 40\,\mathrm{km\,s}^{-1}$ as a minimal proxy for the expected radiatively-driven turbulence \citep[see, e.g.,][]{Moens2022,Moens+2025,Gonzalez-Tora+2025}. In the models shown in this work, we use $T_\ast$, $\dot{M}$, and luminosity $L$ as input parameters. The chemical composition is set to solar abundances except $X_\text{H} = 0.5$, $X_\text{He} = 0.48$, and $X_\text{N} = 0.013$ with depleted C and O ($\log X_\mathrm{C} = -3.8$, $\log X_\mathrm{O} = -3.2$) to account for the expected CNO equilibrium at the surface. Each model then yields the terminal wind velocity and the stellar mass as an output value \citep[see, e.g.,][]{Sander+2020,Bernini-Peron+2025}. Based on our constraints, we computed dedicated PoWR$^\textsc{hd}$ atmospheres for $50 \leq T_\ast/\mathrm{kK} \leq 70$ and $-6.8 \leq \log \dot{M} \leq -6.0$ and fixing $X=0.5$.
After an initial guess for $L$ and a preliminary analysis with shifted models to match the observed SED, the PoWR$^\textsc{hd}$ model sequences were eventually recalculated with the established luminosity to produce our final model set.

\subsection{Parameter inference setup}

We inferred the system parameters by jointly fitting the photometric SED and the normalised optical spectra within a Bayesian framework. The parameter vector is 
$$\vec\theta = \left(d, E(B-V), R_V, T_{\ast,1}, \log g_1, L_1, T_{\ast,2}, \log \dot{M}_2, L_2, s\right)\,,$$
where $d$ is the distance, $E(B-V)$ the colour excess, $R_V$ the total-to-selective extinction ratio, and $L_{1,2}$ the luminosities of the primary and the companion. In practice, the luminosity of the primary is sampled via a multiplicative scaling factor applied to the model spectrum, since the primary atmosphere models can be rescaled at fixed $T_{\ast,1}, \log g_1$, whereas the luminosity of the companion is an explicit parameter of the model grid.

For each parameter set $\vec\theta_i$, model SEDs for both components were interpolated from the PoWR grids, reddened using the \citet{Fitzpatrick1999} extinction law ($A_V = R_V \times E(B-V)$), and scaled by distance. The combined SED was then used to compute synthetic photometry with \texttt{pyphot} \citep{Fouesneau_Lancon2026}.

The synthetic spectra were convolved with instrumental and rotational broadening kernels, Doppler-shifted using the measured epoch RVs, and resampled onto the observed wavelength grids. The component spectra were then combined according to the optical flux ratio implied by the SED. We determined the projected rotational velocity of the primary ($\varv_\mathrm{rot} \sin i$) independently by fitting selected \ion{He}{i} absorption lines (\ion{He}{i\,$\lambda\lambda\, 4471,\,5876,\,7065$}). Best-fit values were obtained via least-squares minimisation, comparing the observed profiles to rotationally broadened model spectra. The resulting $\varv_\mathrm{rot} \sin i$ value was adopted in the composite modelling.

The likelihood combines photometric, spectroscopic, and parallax constraints; $\ln \mathcal{L} =  \ln \mathcal{L}_{\mathrm{spec}}  + \ln \mathcal{L}_{\mathrm{phot}} + \ln \mathcal{L}_\varpi$, where each component was assumed to follow a Gaussian likelihood. 
For the photometric data, the residuals between the observed and synthetic fluxes were weighted by their observational uncertainties. For the astrometric constraint, the model parallax ($1000/d$) was compared to the zero-point-corrected \textit{Gaia} DR3 parallax. The corrected parallax is $\varpi = \varpi_0 - \varpi_\mathrm{ZP} = 0.1485 \pm 0.0153$\,mas, where $\varpi_0$ is the reported \textit{Gaia} DR3 parallax and we adopted the zero-point correction $\varpi_\mathrm{ZP} = -0.0364$\,mas from \citet{Lindegren+2021}. For the spectroscopic data, we introduced an additional spectroscopic jitter term $s$ to account for residual systematic effects not captured by the formal flux uncertainties. The jitter was added in quadrature to the formal uncertainties, resulting in an effective variance $\sigma_{\mathrm{eff},i}^2 = \sigma_i^2 + s^2$ used in the likelihood, and we inferred $s$ simultaneously with the stellar parameters.

We performed posterior sampling with \texttt{UltraNest} using a slice sampler. We adopted a generalised gamma distribution prior on distance, following \citet{Bailer-Jones+2021} and uniform priors within physically motivated bounds for all other parameters.

\subsection{Results}
\label{sec:spectral_fit_results}

\begin{table}[t]
        \caption{Parameters derived for the WR\,2-1 binary system from combined spectroscopic and photometric analysis.}
    \label{tab:stellar_parameters}
        \centering
        \renewcommand{\arraystretch}{1.6}
        \begin{tabular}{lcc}
                \hline
                \hline
                \vspace{0.1cm}
                &       O star &  WR\,2-1 (stripped star) \\
                \hline
                $T_\ast$ (kK) & $36\pm1$ & $60^{+3}_{-2}$ \\
                $T_{2/3}$ (kK)  & $36\pm1$ & $54^{+3}_{-2}$ \\
                $\log g$ (cm\,s$^{-2}$)  & $4.1\pm0.1$ & -- \\
                $\log L$ ($L_\odot$)  & $4.95^{+0.06}_{-0.08}$ & $4.91^{+0.05}_{-0.14}$ \\
                $R_\ast$ ($R_\odot$)  & $7.6^{+0.6}_{-0.7}$ & $2.7^{+0.3}_{-0.5}$ \\
                $\log \dot{M}$ ($M_\odot \mathrm{yr}^{-1}$)\tablefootmark{\dag}  & $-7.8$ & $-6.3 \pm 0.1$ \\
                $\varv_{\infty}$ (km\,s$^{-1}$)\tablefootmark{\dag} & 2933 & 860 \\
                $\varv_\mathrm{rot} \sin i$ (km\,s$^{-1}$)  & $190\pm60$ & -- \\
                $X_{\rm H}$ (mass fr.)  & 0.738 \textit{(fixed)} & 0.5 \textit{(fixed)}\\
                $Y_{\rm He}$ (mass fr.)  & 0.251 \textit{(fixed)} & 0.48 \textit{(fixed)}\\
                $M_\mathrm{spec}$ ($M_\odot$)  & $34_{-6}^{+5}$ & $3.2_{-0.2}^{+0.6}$ \\
                $M_\mathrm{evol}$ ($M_\odot$)  & $24_{-2}^{+1}$ & -- \\
                $M_\mathrm{dyn/evol}$ ($M_\odot$)  & -- & $4.1_{-0.4}^{+0.2}$ \\
                $M_\mathrm{dyn/spec}$ ($M_\odot$)  & -- & $5.8_{-1.1}^{+0.9}$ \\
                $\log\,Q_{\mathrm H}$ (s$^{-1}$)\tablefootmark{\dag} & 48.3 & 48.8 \\
                $\log\,Q_{\mathrm {He\,\textsc{i}}}$ (s$^{-1}$)\tablefootmark{\dag} & 47.1 & 48.4 \\
                $\log\,Q_{\mathrm {He\,\textsc{ii}}}$ (s$^{-1}$)\tablefootmark{\dag}  & 41.7 & 41.2 \\
        \hline
        $d$ (pc) & \multicolumn{2}{c}{$7160^{+330}_{-580}$} \\
        $E_{\mathrm{B-V}}$ (mag) & \multicolumn{2}{c}{$1.68 \pm 0.02$} \\
        $R_V$ & \multicolumn{2}{c}{$3.28\pm0.03$} \\
                \hline
        \end{tabular}
        \tablefoot{$\log\,Q_{\mathrm{H}}$, $\log\,Q_{\mathrm {He\,\textsc{i}}}$, $\log\,Q_{\mathrm {He\,\textsc{ii}}}$ denote the inferred emission rates of H- and He~\textsc{i, ii}-ionising photons. \tablefoottext{\dag} {Parameters without quoted uncertainties were not directly part of the inference but are reported here for the best-fit model.}}
\end{table}

\begin{figure*}
    \centering
    \includegraphics[width=\textwidth]{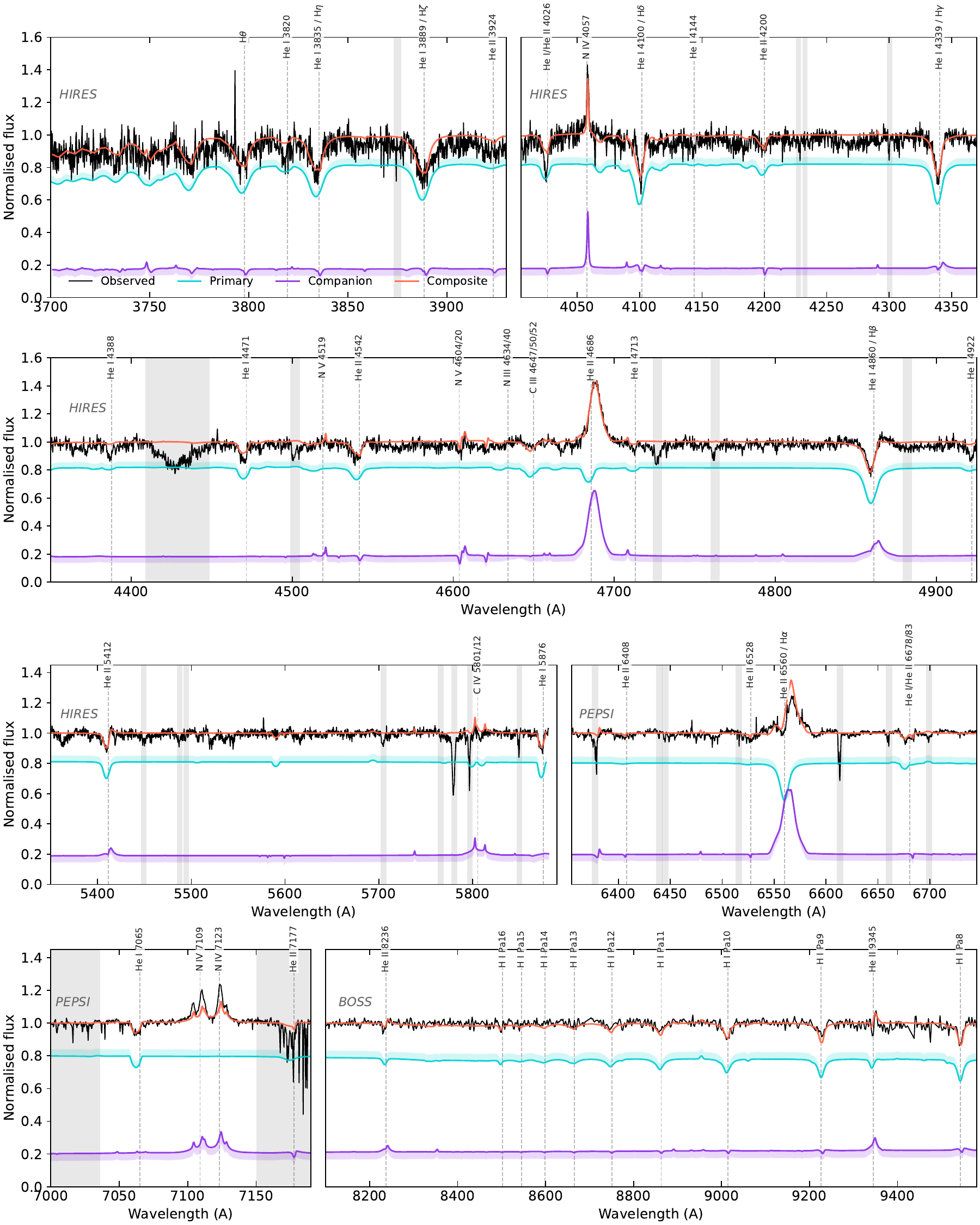}
    \caption{Observed spectra of the WR\,2-1 binary compared to the best-fit binary model. The composite model spectrum (orange) is the sum of the primary O-star spectrum (cyan) and the hot stripped companion star (purple). Shaded bands reflect uncertainties in the fitted spectra, which for the component spectra are dominated by the uncertainty in flux ratio. Narrow deep absorption lines in the observed spectra originate from telluric and interstellar absorption (shaded grey).} 
    \label{fig:composite_fit_detail}
\end{figure*} 

\begin{figure*}
    \centering
    \includegraphics[width=\textwidth]{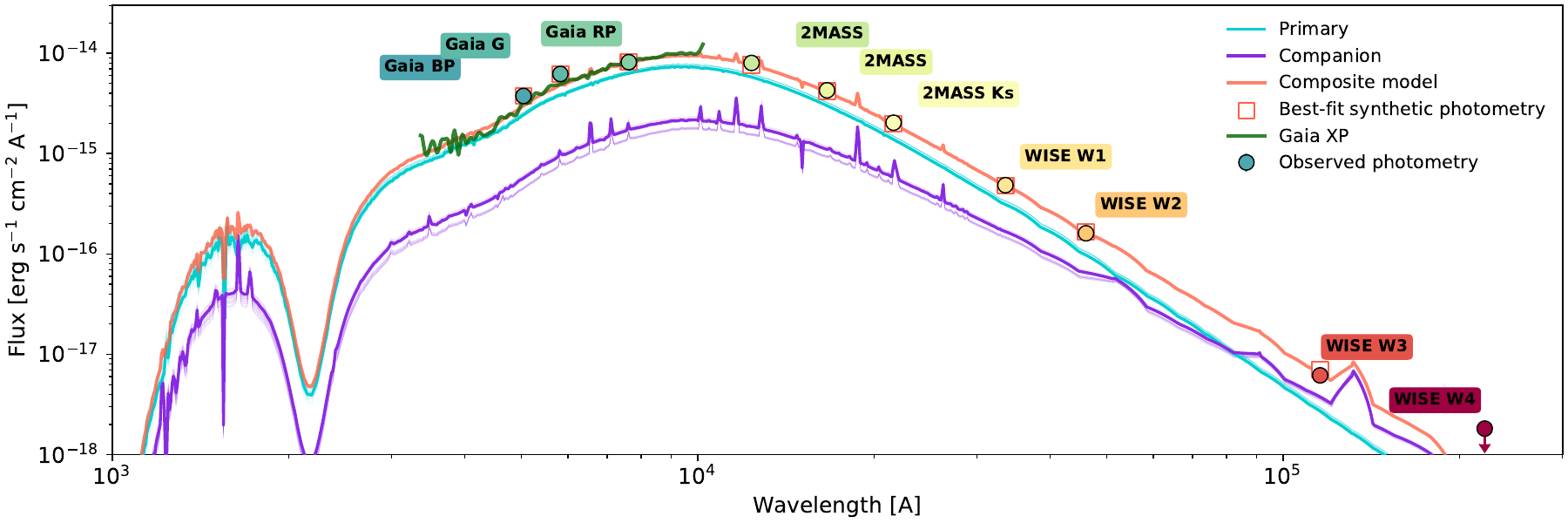}
    \caption{{Comparison of the best-fit composite binary model SED with observed photometry for the WR\,2-1 system}. Coloured markers indicate photometric measurements (see Table~\ref{tab:phot_overview}), orange squares show synthetic photometry generated from the model SED in the same filters. The composite model is shown in orange and was computed as the sum of contributions from the primary O-type star (cyan) and the stripped star companion (purple). For the stellar component spectra, example draws from the posterior are shown as thin lines. For reference, the \textit{Gaia} XP spectrum of the system is overplotted in green.}
    \label{fig:composite_sed_fit}
\end{figure*} 
The resulting best-fit composite model is shown for the optical spectra in Fig.~\ref{fig:composite_fit_detail} and for the SED in Fig.~\ref{fig:composite_sed_fit}. The inferred parameters are summarised in Table~\ref{tab:stellar_parameters}, where we report the median values and the 16th/84th percentile uncertainties from the posterior distributions. These formal uncertainties do not reflect the systematic uncertainties associated with the adopted model atmosphere grid, which we expect to dominate.

We derive a colour excess of $E(B-V) = 1.68\pm 0.02$\,mag and $R_V = 3.28\pm0.03$, corresponding to an extinction $A_V = 5.52 \pm 0.02$\,mag. The high extinction is consistent with the location of the system at a large distance $d = 7160^{+330}_{-580}$\,pc in the outer Galactic disc and with the low observed flux at short wavelengths.

For the primary star, we infer a temperature of $T_{\ast,1} = 36\pm1$\,kK and a surface gravity of $\log g_1 = 4.1\pm0.1$. The inferred bolometric luminosity is $\log L_1/\mathrm{L}_\odot = 4.95^{+0.06}_{-0.08}$, which corresponds to a stellar radius of $R_{\ast,1} = 7.6^{+0.6}_{-0.7}\,R_\odot$. The projected rotational velocity is $\varv_\mathrm{rot} \sin i \approx 190$\,km\,s$^{-1}$.
For the companion, we derive $T_{\ast,2} = 60^{+3}_{-2}$\,kK and a mass loss rate of $\log \dot{M}/(M_\odot\,\mathrm{yr}^{-1}) = -6.3 \pm 0.1$. This uncertainty reflects the formal fitting uncertainty and does not account for variations of the clumping factor. While our models reproduce the observed emission-line wings well, the current spectra do not tightly constrain the degree of clumping. In particular, slightly lower density contrasts, that is, a more smooth wind, cannot be excluded, although a completely smooth wind is unlikely at these temperatures \citep[e.g.,][]{Driessen+2019,Moens2022}. Without UV spectroscopy to constrain the wind structure, we estimate an additional systematic uncertainty in $\dot{M}$ up to $\sim$$0.4\,$dex.
We infer $\log L_2/\mathrm{L}_\odot = 4.91^{+0.05}_{-0.14}$ and a stellar radius of $R_{\ast,2} = 2.7^{+0.3}_{-0.5}\,R_\odot$. The continuum flux contribution of the companion in the optical is approximately 19\%.

\begin{figure}[t]
    \centering
        \includegraphics[width=\columnwidth]{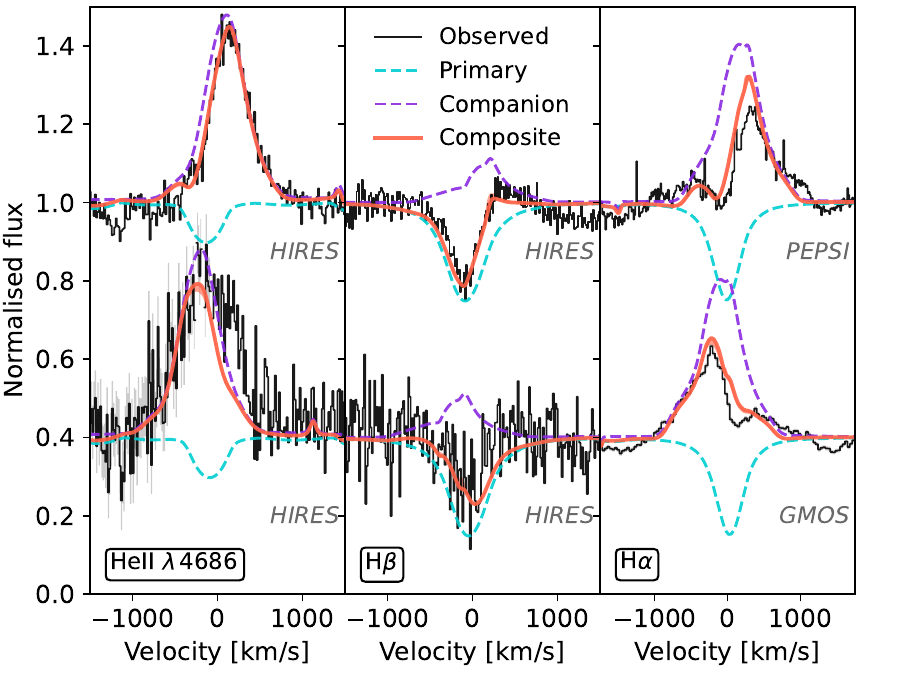}
   \caption{Observed spectra of the WR\,2-1 system near quadrature compared to the best-fit binary model. Two epoch spectra observed near opposite quadratures are shown in black with vertical offsets for readability and zoomed in on the \ion{He}{ii $\lambda \, 4686$} (left), Balmer H$\beta$ (centre), and H$\alpha$ (right) regions. The composite model spectrum (orange) at each epoch is the sum of the Doppler-shifted spectra of the primary O-star (cyan) and the stripped companion star (purple). The RV shifts in the stellar components reproduce well the line profile variations seen in the observed spectra.}
    \label{fig:fit_zoom}
\end{figure}

The composite model reproduces the observed photometry well (Fig.~\ref{fig:composite_sed_fit}), including the flux-calibrated \textit{Gaia} XP spectrum, shown for comparison. Similarly, the optical spectra are well matched (Fig.~\ref{fig:composite_fit_detail}), including the phase-dependent variability of the line profiles (Fig.~\ref{fig:fit_zoom}).
For the companion, the model captures the strength and line shape of the principal emission features, including \ion{He}{ii} and \ion{C}{iv\,$\lambda\lambda\,5801/12$}, as well as the weak P-Cygni profiles in \ion{N}{v\,$\lambda\lambda\,4604/20$}. The \ion{N}{iv} emission lines are slightly under-predicted in strength, which may indicate residual uncertainties in the adopted wind parameters or a higher nitrogen abundance.
With \ion{H}{$\beta$} fully in emission, the companion would be spectroscopically classified as a WR star. We therefore assign it the identifier WR\,2-1. Based on the relative strength of the nitrogen emission lines, its most likely spectral subtype is WN5h \citep{Crowther_Walborn2011}.
For the primary, the hydrogen and most helium lines are well reproduced. However, some \ion{He}{i} lines in the blue (\ion{He}{i\,$\lambda\lambda\,3820,\,4026,\,4388$}) are stronger in the observations than in the model. This discrepancy cannot be resolved by lowering the temperature without degrading the overall fit and may instead point to a modest helium enrichment.
We discuss the implications of the inferred stellar parameters in the context of stripped stars and post-interaction binaries in Sect.~\ref{sec:discussion}.

\subsection{Stellar masses and orbital inclination}
\label{sec:stellar_masses}

\begin{figure}[t]
    \centering
        \includegraphics[width=\columnwidth]{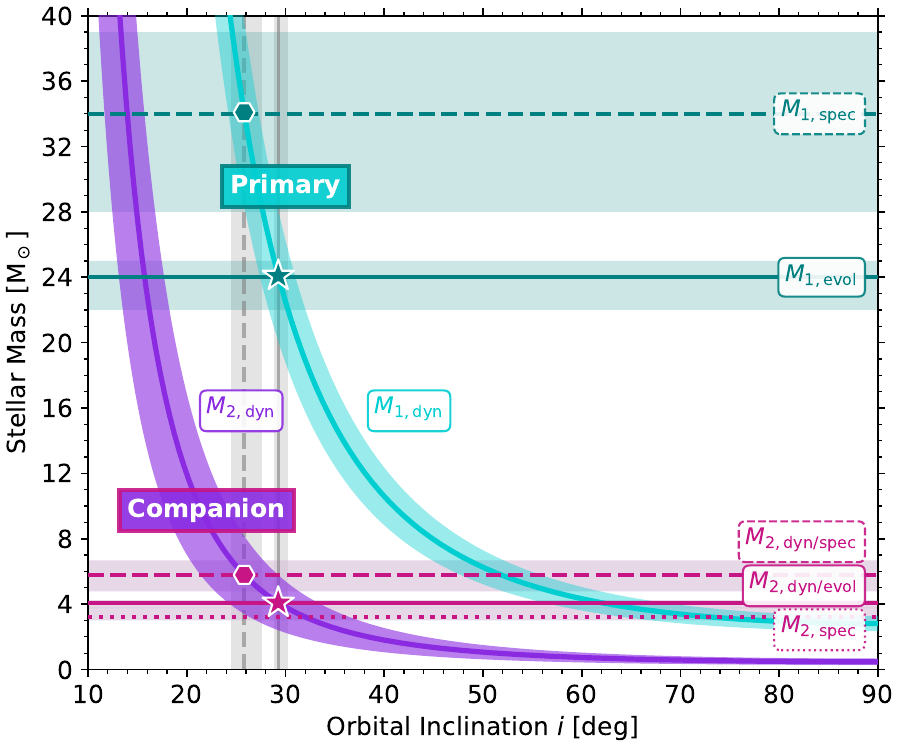}
        \caption{Stellar mass constraints for the WR\,2-1 binary. Dynamical masses derived from the orbital solution are shown as a function of orbital inclination for the primary (cyan) and the companion (purple), with shaded regions indicating $1\sigma$ uncertainties. The evolutionary/spectroscopic mass estimates of the primary are indicated by the horizontal solid/dashed teal lines. The corresponding companion masses $M_\mathrm{2,\,dyn/evol} = 4.1_{-0.4}^{+0.2}\,\mathrm{M}_\odot$ / $M_\mathrm{2,\,dyn/spec} = 5.8_{-1.1}^{+0.9}\,\mathrm{M}_\odot$, inferred from the dynamical mass ratio, are shown as pink solid/dashed lines. The spectroscopic mass of the companion is shown with a pink dotted line. The intersection of these constraints implies an orbital inclination of $i = 26^\circ$--$29^\circ$.}
    \label{fig:mass_constraints}
\end{figure}

We estimated the mass of the primary star using both spectroscopic and evolutionary constraints. Combining the inferred stellar radius and surface gravity yields a spectroscopic mass of $M_\mathrm{1, spec} = 34_{-6}^{+5}\,\mathrm{M}_\odot$, where the uncertainty is dominated by the error in $\log g_1$.
An evolutionary mass was obtained by comparing the stellar parameters of the primary ($L_{1}$, $T_\mathrm{2/3, 1}$) with single-star evolutionary tracks using the \textsc{BONNSAI}\footnote{The \textsc{BONNSAI} web-service is available at \url{www.astro.uni-bonn.de/stars/bonnsai}} tool \citep{Schneider+2014} and the models of \citet{Brott+2011}. We adopted flat priors on mass, age, and initial rotation. This yields an evolutionary mass of $M_\mathrm{1, evol} = 24_{-2}^{+1}\,\mathrm{M}_\odot$, lower than the spectroscopic value, with the two estimates differing at approximately the $2\sigma$ level. 
Although the evolutionary mass has smaller formal uncertainties, it relies on the assumption that the mass gainer can be adequately described by single-star evolutionary models and has largely returned to an equilibrium state following the binary interaction phase. Discrepancies between spectroscopic and evolutionary masses are not a new phenomenon. In the context of single stars, \citet{Herrero+1992} reported evolutionary masses for Galactic OB stars that were systematically higher than those inferred from spectroscopy. A similar effect has also been found for detached binary systems. For example, \citet{Tkachenko+2020} reported a systematic overestimation of evolutionary masses relative to the model-independent dynamical masses of stars. In the present study, however, we find the opposite behaviour: the evolutionary mass is lower than the spectroscopic mass, that is, an inverse manifestation of the classical mass discrepancy.

If we combine the dynamical mass $M_{1}\sin^3 i$ from the orbital solution with the inferred primary mass, that provides a corresponding estimate of the orbital inclination. As illustrated in Fig.~\ref{fig:mass_constraints}, the evolutionary (spectroscopic) mass implies an inclination of $i = 29 \pm 1^\circ$ ($i = 26 \pm 2^\circ$). This corresponds to a de-projected rotational velocity of the primary of approximately $\varv_\mathrm{rot} \simeq 390\,\mathrm{km\,s}^{-1}$ ($430\,\mathrm{km\,s}^{-1}$) and a rotational period of the star of $P_\mathrm{rot} \simeq 1.0\,$d, that is, shorter than the orbital period. Such rapid, supersynchronous but subcritical rotation is commonly observed in WR+O binaries, where it is generally attributed to spin-up during Roche-lobe overflow \citep{Shara+2017,Shara+2020}. We emphasise that this does not constitute an independent inclination measurement, but rather the inclination implied by the adopted primary mass.

For the companion, the dynamical mass $M_2\sin^3 i = 0.48_{-0.18}^{+0.21}\,\mathrm{M}_\odot$ together with the evolutionary (spectroscopic) inclination estimate yields a mass of $M_\mathrm{2,\,dyn/evol} = 4.1_{-0.4}^{+0.2}\,\mathrm{M}_\odot$ ($M_\mathrm{2,\,dyn/spec} = 5.8_{-1.1}^{+0.9}\,\mathrm{M}_\odot$). An independent constraint on the companion mass comes from the best-fitting PoWR$^\textsc{hd}$ models, which have masses $M_{2,\mathrm{spec}} \simeq 3.2_{-0.2}^{+0.6}\,\mathrm{M}_\odot$. These estimates consistently place the companion within the frequently quoted $2$--$8\,\mathrm{M}_\odot$ mass range of intermediate-mass stripped stars and at the low end of what has been inferred for (massive) WR stars. Recently, the analysis of WR\,58 with PoWR$^\textsc{hd}$ yielded a mass of $4.9\pm0.8\,M_\odot$ \citep{Lefever+2026}. Now WR\,2-1 appears to be among the lowest-mass WR-like objects currently known with dynamically consistent atmosphere modelling.

\section{Discussion}
\label{sec:discussion}

\subsection{Post-interaction nature of the system}

The observed and inferred properties of the WR\,2-1 binary point to a system composed of two stars with markedly different temperatures. The spectra exhibit photospheric absorption lines characteristic of a fast-rotating O-type main-sequence star and high-ionisation emission lines indicative of a significantly hotter component. The RV variability of these features implies a short orbital period ($P \sim 5.9\,$d) and a large mass ratio ($q \sim 5.8$), with the hotter component being the less massive object. This configuration is characteristic of a post-interaction system in which the initially more massive and hotter star was stripped through binary interaction, the initially less massive star gained mass and angular momentum, and the system underwent a mass-ratio reversal.

The composite spectroscopic and photometric modelling yields a best-fit model of the system featuring a $\sim 36$\,kK O-type primary and a $\sim 60$\,kK helium-star companion. The inferred parameters of the companion place it bluewards of the zero-age main sequence (ZAMS), a region not populated by single-star evolutionary tracks but expected for envelope-stripped stars. The inferred mass of $M_2 = 3.2$--$5.8\,\mathrm{M}_\odot$ places it squarely in the intermediate-mass regime ($\sim 2$--8\,$M_\odot$).

The available evidence strongly disfavours several alternative interpretations of the system. In particular, the emission-line star is unlikely to have formed through single-star evolution by shedding its hydrogen envelope through stellar winds alone. Such a scenario would require the progenitor to have been initially more massive than the present-day O-star companion in order to drive sufficiently strong winds. Even then, such stars typically remain substantially more massive than the $3.2$--$5.8$\,M$_\odot$ companion inferred here and exhibit considerably stronger winds ($\log \dot{M}/(M_\odot\,\mathrm{yr}^{-1}) \gtrsim -5$). If strong intrinsic mass loss had occurred during a cooler evolutionary stage (e.g., as a red supergiant), it would have been hard to avoid mass transfer given the short orbital period. Finally, the combination of a low-mass WR-like star, a massive ($\simeq 24$--34\,M$_\odot$) and only mildly evolved O-star companion, and the extreme mass ratio is naturally explained by Roche-lobe overflow and mass transfer, but is difficult to reproduce through the independent evolution of the two stars.
A non-stellar origin of the emission features is also unlikely: while lines such as H$\alpha$ and \ion{He}{ii\,$\lambda4686$} are commonly observed in accretion discs \citep{Steeghs_Casares2002, Halpern+2018, Koljonen+2023}, the presence of strong \ion{N}{iv} emission is not expected for accreting compact objects. The P-Cygni-shaped line profiles (e.g. in \ion{N}{iv} and \ion{N}{v}) further support a stellar wind origin.

We therefore interpret WR\,2-1 as a post-interaction binary consisting of a hot intermediate-mass stripped star -- the former mass donor -- and a now more massive O-type star that has accreted material during binary interaction.

\subsection{WR\,2-1 in the context of known stripped stars}
The well-constrained parameters of WR\,2-1 provide an opportunity to place this system in the growing population of detected stripped stars and to compare its properties with theoretical predictions.

\subsubsection{Hertzsprung-Russell diagram}
\begin{figure}[t]
    \centering
    \includegraphics[width=\columnwidth]{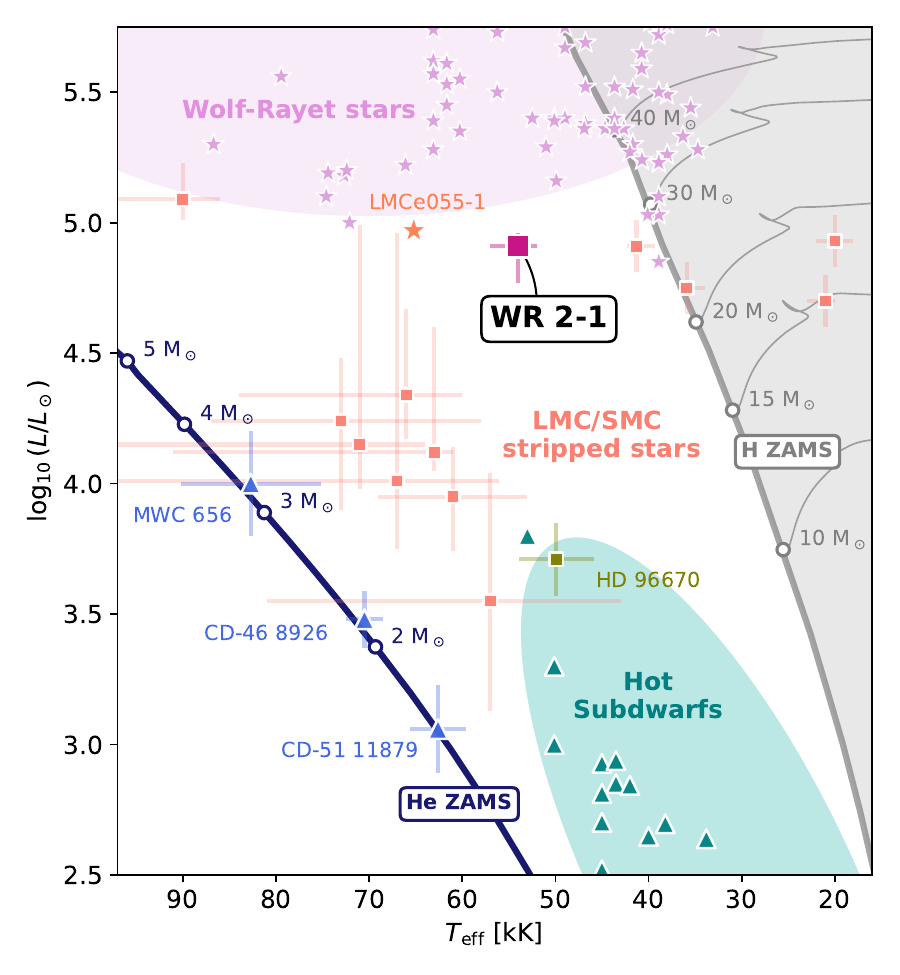}
    \caption{Hertzsprung-Russell diagram showing the position of WR\,2-1 (pink) compared to known envelope-stripped stars. The figure includes Galactic WR stars (light pink), hot subdwarfs (teal), with the subset of the hottest subdwarfs and merger candidates highlighted in blue, stripped stars in the Magellanic Clouds (orange), and the candidate intermediate-mass stripped star HD~96670 (olive); see main text for references. Its inferred temperature and luminosity place WR\,2-1 in the intermediate regime between hot subdwarfs and classical WR stars.}
    \label{fig:lit_comparison_hrd}
\end{figure}

In Fig.~\ref{fig:lit_comparison_hrd}, we compare the position of WR\,2-1 in the Hertzsprung-Russell diagram with known helium stars and stripped-star candidates. WR\,2-1 occupies the region between the hydrogen and helium ZAMS, where stripped stars are expected to reside. The helium ZAMS shown in the figure is from \citet{Picco+2024} for solar metallicity, hydrogen ZAMS and reference single-star evolutionary tracks are taken from the MIST library \citep{Dotter2016, Choi+2016}.

WR\,2-1 lies in the gap between the populations of hot subdwarfs and classical WR stars. It is substantially more luminous than hot subdwarfs \citep[e.g.][]{Wang+2021,Peters+2013,Peters+2016,Mourard+2015,Klement+2022}, including also the hotter systems CD–46~8926 and CD–51~11879 \citep{Krticka+2024} and the Be-binary MWC~656 \citep{Mueller-Horn+2026}, while lying at the very low-luminosity end of the Galactic WR-star distribution \citep{Sander+2012,Sander+2019,Hamann+2006,Hamann+2019, Zhang+2020, Todt+2010, Lefever+2026}.

The closest analogues are found among the growing sample of intermediate-mass stripped stars identified in the Magellanic Clouds. These include the systems of \citet{Drout+2023} and \citet{Goetberg+2023}, discovered through UV excess, the partially stripped SMC binaries analysed by \citet{Ramachandran+2023,Ramachandran+2024}, and the low-luminosity WR star LMCe055-1 \citep{Massey+2024}. Together, these objects populate the region bracketed by the hot subdwarf and WR regimes, with luminosities of $\sim10^4$--$10^5\,L_\odot$ and temperatures typically above 50\,kK (except for the four partially stripped SMC stars). WR\,2-1 fits naturally within this population.
Among Galactic systems, the closest currently known analogue is HD~96670 \citep{Naze_Rauw2025}, whose inferred mass is similar to that of WR\,2-1. However, HD~96670 is cooler and approximately an order of magnitude less luminous. It also remains less constrained spectroscopically, and UV follow-up spectroscopy will be required to establish its wind properties.

\subsubsection{Wind mass-loss rates}
\begin{figure}[t]
    \centering
    \begin{minipage}[t]{\columnwidth}
        \centering
        \includegraphics[width=\columnwidth]{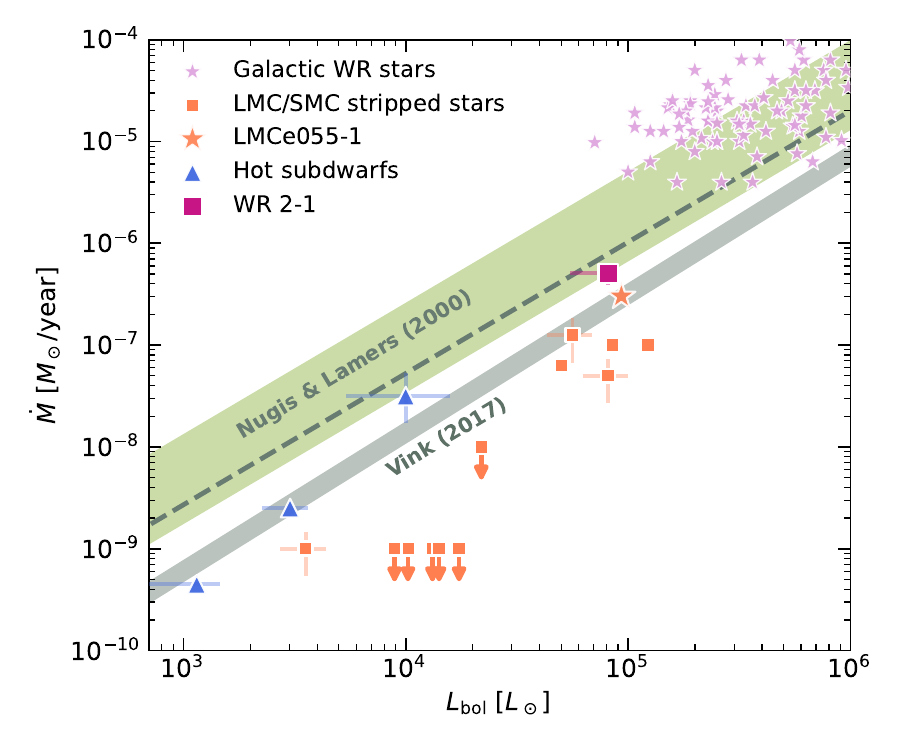}
    \end{minipage}
    \begin{minipage}[t]{\columnwidth}
        \centering
        \includegraphics[width=\columnwidth]{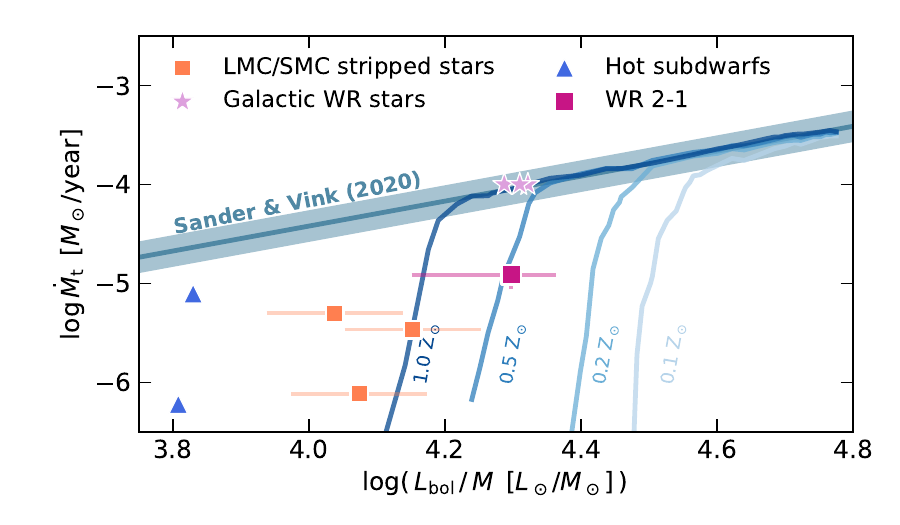}
    \end{minipage}
    \caption{Wind parameters WR\,2-1 and comparison samples. \textit{Top:} Mass-loss rate as a function of luminosity; colour coding follows Fig.~\ref{fig:lit_comparison_hrd}. The green and grey shaded regions indicate theoretical predictions from the empirical WR relation of \citet{Nugis_Lamers2000} and the stripped-star prescription of \citet{Vink2017}, respectively. The dashed line shows the relation for $Y=0.5$. WR\,2-1 occupies the intermediate regime between hot subdwarfs and classical WR stars, consistent with \citeauthor{Nugis_Lamers2000}. \textit{Bottom:} Transformed mass loss rate vs. luminosity-to-mass ratio for stripped stars with dynamically consistent mass estimates. Here, we assumed $M_\mathrm{2, \,dyn/evol}$ to compute the luminosity-to-mass ratio -- using $M_\mathrm{2, \,dyn/spec}$ or $M_\mathrm{2, \,spec}$ changes the value by $\sim\pm0.15$. WR\,2-1 falls below the inferred relation for helium star winds by \citet{SanderVink2020}.} 
    \label{fig:lit_comparison_mdot}
\end{figure}

In Fig.~\ref{fig:lit_comparison_mdot}, we compare the inferred mass-loss rate of WR\,2-1, $\log \dot{M}/(M_\odot\,\mathrm{yr}^{-1}) = -6.3 \pm 0.1$, to other stripped stars. The combined sample suggests a continuous increase of mass-loss rate with luminosity, with WR\,2-1 lying in the intermediate regime between hot subdwarfs and classical WR stars, consistent with its position in the Hertzsprung-Russell diagram.

The derived mass-loss rate is in good agreement with the empirical WR relation of \citet{Nugis_Lamers2000}, but is a factor of $\sim2$--$3$ above predictions from the helium-star wind models of \citet{Vink2017}, which generally yield lower mass-loss rates in this regime.
This discrepancy could indicate that current helium-star prescriptions underestimate mass loss in the intermediate-mass regime. However, WR\,2-1 was identified through its emission-line signature, introducing a bias towards systems with relatively strong winds, and the result may therefore not be representative of the underlying population. 
At fixed luminosity, Galactic stripped stars exhibit systematically higher mass-loss rates than their counterparts in the Magellanic Clouds, consistent with the expected metallicity dependence of radiatively driven winds.

The lower panel of Fig.~\ref{fig:lit_comparison_mdot} shows the transformed mass-loss rate $\dot{M}_\mathrm{t}$ as defined by \citet{Graefener_Vink2013}. This quantity rescales $\dot{M}$ to a reference luminosity, terminal velocity, and clumping factor, enabling more direct comparison between stars of different luminosities. \citet{SanderVink2020} predict a linear relation between $\dot{M}_\mathrm{t}$ and the luminosity-to-mass ratio for optically thick WR winds, with a breakdown expected when the winds become optically thin. While low-luminosity WR stars closely follow this relation \citep{Lefever+2025}, WR\,2-1 falls below the trend, indicating a transition into the optically thin wind regime.

The stellar wind of the stripped companion may produce thermal free--free radio emission, while interaction between the winds of the two stars could additionally give rise to non-thermal radio and X-ray emission, as observed in other colliding-wind binaries \citep[e.g.,][]{1986ApJ...303..239A,1987ApJ...320..283P}. Future radio observations with facilities such as the Square Kilometre Array (SKA) or the next-generation Very Large Array (ngVLA) could provide valuable additional constraints on the wind properties of WR\,2-1.

\subsubsection{Comparison with low-luminosity WR stars}
The transition from hot subdwarfs through intermediate-mass stripped stars to classical WR stars is likely continuous rather than marked by a sharp physical boundary. In this context, WR\,2-1 occupies an interesting position at the interface between the intermediate-mass stripped-star and WR populations. Its temperature and luminosity are comparable to those of Galactic low-luminosity WR stars, although it lies at the low-luminosity end of the distribution (Fig.~\ref{fig:lit_comparison_hrd}). 
The inferred mass of WR\,2-1 lies below those reported for Galactic WR stars, but only by a modest margin relative to recent examples such as WR\,58 \citep[$4.9\pm0.8,M_\odot$;][]{Lefever+2026}.
The most notable distinction with classical WR stars is the wind strength. WR\,2-1 has a mass-loss rate approximately an order of magnitude lower than those typically inferred for Galactic low-luminosity WR stars (Fig.~\ref{fig:lit_comparison_mdot}).
A particularly interesting comparison object is LMCe055-1 \citep{Massey+2024}, a low-luminosity WR star in the LMC, whose temperature, luminosity, and comparatively weak wind closely resemble those of WR\,2-1. Interestingly, despite being part of a close binary system, \citet{Massey+2024} conclude that LMCe055-1 did not form through binary stripping.

\subsection{Surface abundances and metallicity}
A detailed abundance analysis is beyond the scope of this work, given the moderate S/N of the spectra and the lack of spectrally disentangled components. We therefore limit the discussion to qualitative constraints from the composite modelling.

The most robust result is that the stripped companion retains a substantial amount of hydrogen and has not been fully stripped to its helium core. This conclusion is primarily based on the strength of the Balmer emission lines, particularly H$\alpha$ and H$\beta$ (see also Appendix~\ref{appendix:hydrogen_frac}). Models with a hydrogen mass fraction of $X = 0.5$ provide a good match to the observed spectra,(Fig.~\ref{fig:composite_fit_detail}), but a precise quantitative determination of the hydrogen abundance will additional high-quality data together with dedicated abundance modelling.

The metal lines provide only limited constraints on chemical enrichment. The observed \ion{C}{iv\,$\lambda\lambda\,5801/12$} lines are consistent with the assumed subsolar carbon abundance ($\log X_C = -3.8$).
For the primary star, no metal lines with sufficient S/N are available for a meaningful abundance analysis. However, the \ion{He}{i} lines in the blue part of the spectrum (e.g. \ion{He}{i\,$\lambda\lambda\,3820,\,4026,\,4388$}) are systematically stronger in the observations than in the best-fit model. Lowering the effective temperature does not resolve this discrepancy without degrading the overall fit, suggesting a modest helium enrichment. Such an enrichment would be consistent with past accretion of helium-rich material during binary mass transfer \citep{Jin+2026}.

The PoWR atmosphere models and binary-evolution grid adopted in this work assume solar metallicity and solar initial abundances. From the inferred distance and sky position, the system is located at a Galactocentric radius of approximately 13.6\,kpc. Abundance tracers in the outer Galactic disc indicate oxygen abundances that are lower than solar by approximately 0.2--0.3\,dex at these radii \citep{Martinez-Hernandez+2026}, suggesting that the progenitor binary may have formed with a moderately subsolar metallicity. This is qualitatively supported by the presence of the nearby WN2 star WR\,2 \citep{Sander+2026} and the position of WR\,2-1 in the lower panel of Fig.\,\ref{fig:lit_comparison_mdot}. 
If in particular also the iron abundance were subsolar, the use of solar-metallicity models could introduce biases in the inferred stellar and evolutionary parameters. In the PoWR$^\textsc{hd}$ atmosphere modelling, a somewhat lower stellar mass would be required to yield the same emission-line spectrum \citep[cf.][]{Sander+2020}. 
In the evolution models, lower metallicity would reduce the strength of radiatively driven winds, particularly also following mass transfer, potentially leading to somewhat different stripped-star properties, including a higher remaining surface hydrogen fraction \citep{Klencki+2022}. However, the winds of post-interaction stars remain poorly constrained and may already be overestimated in current evolutionary calculations. Quantifying the effects of subsolar metallicity therefore requires dedicated atmosphere and binary-evolution models, which we defer to future work.

\subsection{Evolutionary history}
\label{sec:evolution}

\begin{figure*}[ht!]
    \centering
    \includegraphics[width=\textwidth]{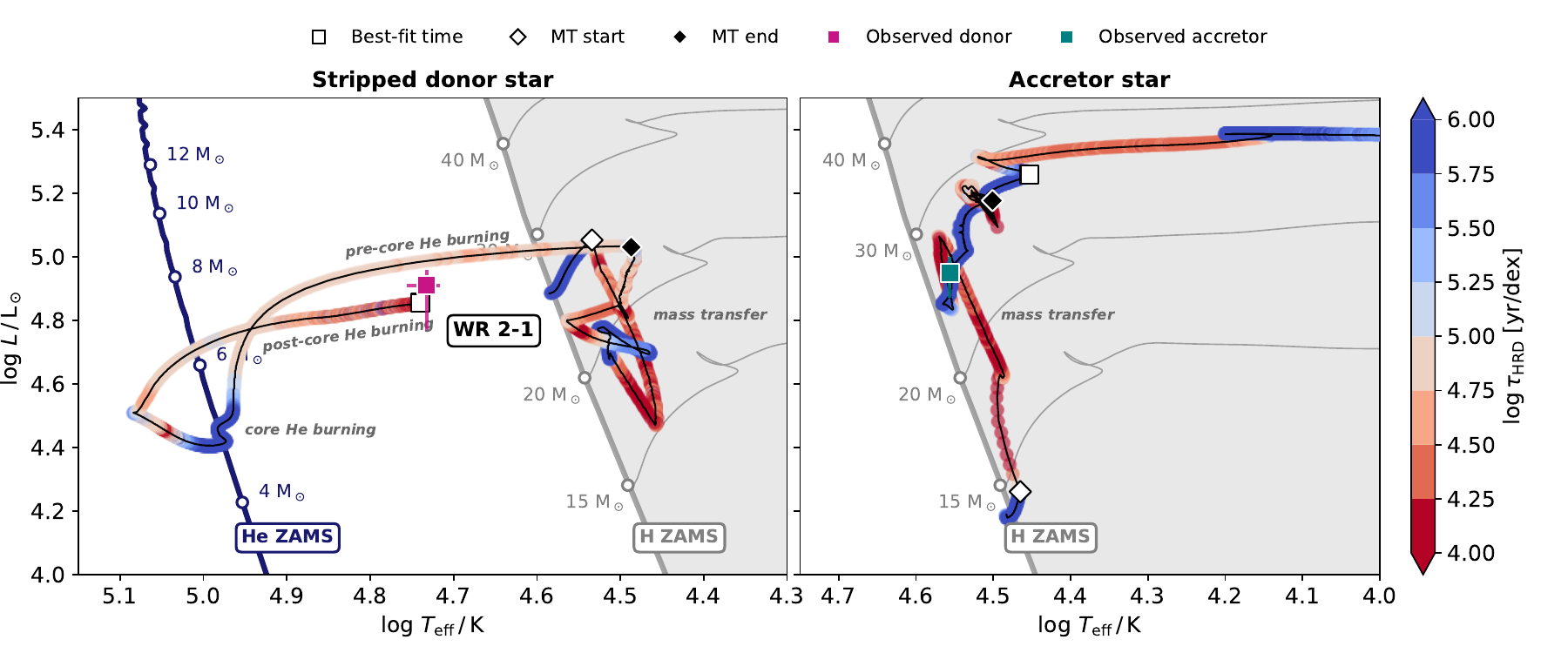}
    \caption{Evolutionary tracks of binary models in the Hertzsprung-Russell diagram for comparison with WR\,2-1. The observed properties of the stripped donor (left) and the mass gainer (right) are compared to a representative model from the grid of \citet{Jin+2026} (black). The tracks are coloured by their ``HRD-timescale'' ($\tau_\mathrm{HRD} = \Delta \nicefrac{t_\mathrm{age}}{\sqrt{[\Delta \log (T_\mathrm{eff} / \mathrm{K})]^2+[\Delta \log (L / \mathrm{L}_\odot)]^2}}$), to illustrate the relative time spent in different evolutionary phases (blue means longer-lived). The onset and end of mass transfer are marked by diamond symbols. The epoch that best matches the observed properties is indicated by white square markers. Hydrogen and helium ZAMS, as well as single-star evolutionary tracks are shown for reference.
    Comparison with the model tracks suggests that the stripped donor star presently is in a transitional stage after mass transfer with a cooler temperature, and higher luminosity compared to the contracted core-helium burning phase.}
    \label{fig:mesa_model_HRD}
\end{figure*}

In order to constrain the evolutionary state of the system, as well as the properties of its progenitor and interaction history, we compared the observed parameters with predictions from binary evolution models.
We scanned the grid of models by \citet{Jin+2026} for systems that reproduce the present-day properties of the WR\,2-1 binary. The grid was computed with the Modules for Experiments in Stellar Astrophysics code \citep[MESA;][]{Paxton+2011, Paxton+2013, Paxton+2015, Paxton+2018, Paxton+2019} and follows the evolution of massive binary systems from the ZAMS to core carbon or helium depletion unless other termination conditions such as the onset of unstable mass transfer are met, with a focus on the post-mass transfer properties and abundances of the stars. The models assume solar metallicity and span initial primary masses of 5--100\,M$_\odot$, mass ratios of 0.1--0.95, and orbital periods of 0.3--5011\,d. The mass transfer efficiency is computed self-consistently; regulated by rotation-limited accretion \citep{Petrovic+2005}.
From the $\sim$38\,000 models in the grid, we selected systems that match the observed properties of both components at some point during their evolution. For the stripped star, we required an approximate match in mass (2--8\,M$_\odot$), luminosity ($4.6 < \log L/\mathrm{L}_\odot < 5.2$), and effective temperature (50--75\,kK). For the accretor, we imposed 20--40\,M$_\odot$, $4.6 < \log L/\mathrm{L}_\odot < 5.4$, and 30--45\,kK. We further required comparable orbital configurations, restricting the sample to systems with $P < 20\,$d and mass ratios $q > 4$, and limited the initial total mass to $20 < (M_\mathrm{1, ini} + M_\mathrm{2, ini})/\mathrm{M}_\odot < 80$.

\subsubsection{Progenitor system}
Applying these criteria identifies 14 models that reproduce the observed stellar parameters and binary configuration of WR\,2-1 at some stage of their evolution (see Figures\,\ref{fig:mesa_model_HRD} and \ref{fig:mesa_model_time}). The corresponding initial conditions occupy a relatively narrow region of parameter space, with donor initial masses of 22--28\,M$_\odot$, companion initial masses of 13--19\,M$_\odot$, initial mass ratios of 0.50--0.85, and orbital periods of 2--4\,d.
The restricted period range reflects the requirement for Case~A mass transfer, which is needed to reproduce the observed properties of the system and occurs only for such short initial periods. Systems with longer initial periods evolve through Case~B mass transfer, while shorter-period systems are expected to merge \citep[][their Fig.~A.2]{Jin+2026}. In the Case~A scenario, mass transfer begins while the donor is still on the main sequence, before the helium core is fully developed, leading to lower-mass stripped stars for a given initial mass. The models yield helium-star masses of $\sim3$--$5\,\mathrm{M}_\odot$ for progenitors above $20\,\mathrm{M}_\odot$, whereas similar helium-star masses would result from Case~B evolution for significantly lower initial masses ($\sim$12--16\,M$_\odot$; \citealt{Goetberg+2018}). At the same time, efficient mass transfer allows the companion to accrete a substantial fraction of the stripped envelope, naturally producing large present-day mass ratios.

\subsubsection{Mass transfer history}
Figure~\ref{fig:mesa_model_HRD} shows the evolutionary tracks of a selected model in the Hertzsprung-Russell diagram. As an illustrative example, we highlight a model with an initial period of 2\,d and a mass ratio of $q_\mathrm{ini} = 0.55$, which provides a close match to the observed system.
In this model, the initially $25\,\mathrm{M}_\odot$ donor fills its Roche lobe after $\sim4$\,Myr, while still on the main sequence. Mass transfer proceeds in multiple episodes over a total duration of $\sim5$\,Myr, during which the donor loses $\sim20\,\mathrm{M}_\odot$ of its hydrogen-rich envelope and is reduced to a mass of $\sim5\,\mathrm{M}_\odot$. The companion accretes $\sim11\,\mathrm{M}_\odot$, increasing its mass to $\sim25\,\mathrm{M}_\odot$.
Following envelope stripping, the donor contracts and heats up to become a compact helium star. This contraction phase lasts for $\sim0.17$\,Myr and is followed by a core-helium burning phase. Subsequently, the star expands again, evolving towards higher luminosities and cooler temperatures over $\sim0.07$\,Myr before reaching core carbon depletion. By this stage, the accretor has evolved off the main sequence. In the following section, we discuss what is the evolutionary phase we currently observe WR\,2-1 in.

\subsubsection{Pre- or post-core helium burning?}
The lifetime of the stripped helium stars shown in Fig.~\ref{fig:mesa_model_HRD} is typically of the order $\sim$1\,Myr, corresponding to about 10\% of the total stellar lifetime. Most of this time ($\sim$80--90\%) is spent in a compact core-helium-burning phase. This phase is preceded and followed by short-lived inflated stages: an initial contraction phase after mass transfer with hydrogen shell burning, and a later expansion phase following core-helium depletion with helium shell burning. During both phases, the star is cooler and more luminous.
The location of WR\,2-1 in the Hertzsprung-Russell diagram suggests that it is unlikely to be observed during core-helium burning. Instead, it appears more consistent with one of the inflated phases. However, the distinction between the contracting and expanding stages is less straightforward.
The contraction phase is longer-lived by approximately a factor of three, making it statistically more likely to be observed. The relatively high surface hydrogen abundance inferred for the stripped star is more naturally explained shortly after envelope stripping, before continued wind mass loss predicted in the MESA models further erodes the residual hydrogen-rich envelope. Additionally, the rapid rotation of the mass gainer may indicate a relatively recent mass transfer.
Conversely, the observed luminosity and inferred mass of the stripped star are reproduced more readily by post-core-helium-burning models. The unusually high luminosity-to-mass ratio, $\log (L/M \ [\mathrm{L}_\odot/\mathrm{M}_\odot]) \simeq 4.3$, also favours a more advanced evolutionary state. In particular, the luminosity exceeds the maximum values predicted for partially stripped, hydrogen-shell-burning stars of comparable mass by \citet{Sabhahit2025}, suggesting that helium-shell burning may contribute as an additional luminosity source.
At this point, we cannot definitively distinguish between the scenarios.

\subsubsection{Future evolution}
The stripped star WR\,2-1 is expected, under current evolutionary models, to undergo core collapse and likely explode as a stripped-envelope supernova \citep{Eldrige+2013, Tauris+2015, Chanlaridis+2022}. Given the presence of residual hydrogen in its envelope, the supernova would likely be of Type~IIb. For helium-core masses in the range $\sim3$--$6\,\mathrm{M}_\odot$, the resulting carbon-oxygen core is expected to be $\sim2$--$4\,\mathrm{M}_\odot$ \citep{Laplace+2021}, suggesting the formation of a neutron star remnant \citep{Maltsev+2025}.
The future evolution of the companion is less certain. With a current mass of $\sim24\,\mathrm{M}_\odot$, it is also expected to end its life in core collapse. Given the short orbital period, the binary is likely to interact again. This may include a phase of reverse mass transfer when the accretor evolves off the main sequence (potentially making it a high mass X-ray binary), or additional stripping of the donor. The system may therefore undergo complex binary evolution prior to the first supernova. 
If the system survives both the supernova explosion and subsequent binary interactions without disruption or merger, it may evolve into a double neutron star system \citep{Tauris+2017, Kruckow+2018}. However, robust predictions of the final outcome require dedicated binary evolution modelling and are beyond the scope of this work.

\subsection{Occurrence rate and selection function}
The discovery of WR\,2-1 in a short-lived inflated phase rather than during core-helium burning is not surprising. As a compact helium star, the stripped star would be substantially hotter ($\sim$$100$\,kK) and less luminous, reducing its optical continuum contribution by roughly an order of magnitude. At the same time, its likely weaker wind would produce much weaker emission lines. In such a state, the stripped star would therefore be considerably more difficult to identify in an optical survey. Our search is thus naturally biased towards inflated stripped stars with stronger winds.

Is the discovery of only one such system in the SDSS-V sample then unexpected? A detailed treatment of the survey selection function is beyond the scope of this work, but a simple order-of-magnitude estimate suggests that it is not. The \texttt{mwm\_ob} sample observed and analysed so far contains approximately $N_\mathrm{OB}\simeq10^3$ OB stars with masses above $\gtrsim20\,M_\odot$ (Rix et al. in prep.). Assuming that $\sim33\%$ of these undergo envelope stripping \citep{Sana+2012} and that the inflated stripped-star phase lasts roughly 1\% of the system lifetime yields an expected value of order three such systems. This estimate does not yet account for search incompleteness or the requirement of sufficiently strong wind emission lines to be recognised. Given these additional biases, the detection of a single system so far is consistent with expectations.
A systematic search for further stripped-star binaries in the SDSS-V sample is ongoing.

\section{Summary and conclusions}
\label{sec:conclusion}

We have identified and characterised the WR\,2-1 system as a Galactic post-interaction binary hosting an intermediate-mass stripped star. The system was discovered in data from the SDSS-V MWM survey and selected for follow-up due to clear signatures of two distinct temperature components. In particular, the presence of high-ionisation emission lines such as \ion{He}{ii} and \ion{N}{iv}, alongside photospheric absorption typical of a cooler O-type star, and the large-amplitude RV variability revealed its binary nature. 

A consistent physical picture emerges from the joint spectroscopic and photometric analysis. The system consists of a rapidly rotating O-type star ($M_\mathrm{1,evol} = 24_{-2}^{+1}\,\mathrm{M}_\odot$, $T_{\ast,1} = 36\pm1\mathrm{kK}$), interpreted as the mass gainer, and a hot companion ($M_2 = 3.2$--$5.8\,\mathrm{M}_\odot$, $T_{\ast,2} = 60^{+3}_{-2}\,\mathrm{kK}$), identified as the stripped remnant of the initially more massive star.
The large mass ratio ($q_\mathrm{dyn}=5.8_{-1.5}^{+2.8}$) derived from the orbital solution, together with the fact that the hotter component is the less massive one, indicates a post-mass transfer configuration where the initial mass ratio has been reversed. This interpretation is supported by the circular short-period orbit ($P=5.944_{-0.007}^{+0.009}$\,d), and the rapid rotation of the mass gainer ($\varv_\mathrm{rot,1} \simeq 390\,\mathrm{km\,s}^{-1}$), both likely consequences of binary interaction. The stripped star retains a significant surface hydrogen fraction ($X \approx 0.5$), indicating only partial envelope stripping.

The position of the stripped star in the Hertzsprung–Russell diagram -- blueward of the main sequence and intermediate between hot subdwarfs and WR stars -- together with its inferred mass and mass-loss rate ($\log \dot{M}_2 = -6.3 \pm 0.1$), supports its classification as an intermediate-mass stripped star. The wind strength is consistent with an extrapolation of empirical WR relations and somewhat higher than current theoretical prescriptions for helium stars, as well as estimates for stripped stars in the Magellanic Clouds. However, we note that our constraints rely on optical diagnostics and that the emission-line-based selection introduces a bias towards systems with relatively strong winds.

Comparison with binary evolution models indicates that the observed properties can be reproduced by progenitor systems with donor initial masses of $\sim$22--28\,M$_\odot$, initial mass ratios of $\sim$0.50--0.85, and orbital periods of $\sim$2--4\,d that undergo Case~A mass transfer with efficient accretion. The stripped star is likely observed in a short-lived inflated phase outside core-helium burning, consistent with its high luminosity ($\log L_2/\mathrm{L}_\odot = 4.91^{+0.05}_{-0.14}$) and comparatively large radius ($R_2 =2.7^{+0.3}_{-0.5}\,\mathrm{R}_\odot$) for its inferred mass. The available constraints do not allow a definitive distinction between a contracting pre- and an expanding post-core-helium-burning phase.
Given its present-day mass, the stripped star is expected to undergo core collapse and likely explode as a stripped-envelope supernova. The system may experience further binary interaction prior to this event and, if it remains bound past the second supernova, could evolve into a double neutron star system. However, robust predictions of its final fate require dedicated binary evolution modelling.

WR\,2-1 provides an important benchmark in the intermediate-mass regime of stripped helium stars, which so far has remained poorly constrained observationally. Its well-determined stellar and binary parameters offer valuable constraints on mass-loss, the efficiency of binary mass transfer, and the structure of partially stripped stars and their accreting companions. As such, the system informs models of stripped-envelope supernova progenitors and compact-object formation channels.
Further progress will require high-resolution, high S/N spectroscopic follow-up to enable spectral disentangling and quantitative abundance analysis. UV spectroscopy would be particularly valuable for constraining wind properties, although this remains observationally challenging given the faintness and high extinction of the system. 

The discovery of WR\,2-1 demonstrates the power of large spectroscopic surveys such as SDSS-V to uncover rare post-interaction systems. Continued exploration of these data, combined with careful modelling of selection effects, will enable a more complete census of stripped stars in the Milky Way and improve our understanding of their role in binary evolution and compact-object formation.

\begin{acknowledgements}
The authors thank Myles Sherman, Geoffrey Moe, Soumyadeep Bhattacharjee, and Cheyanne Shariat for their help in obtaining follow-up data.
This research was supported in part by grant NSF PHY-2309135 to the Kavli Institute for Theoretical Physics (KITP). This research benefited from discussions at the 
''Stellar-Mass Black Holes at the Nexus of Optical, X-ray, and Gravitational Wave Surveys'' programme at the Kavli Institute of Theoretical Physics. J.M.-H. and H.-W.R. acknowledge support from the European Research Council for the ERC Advanced Grant [101054731]. This work was supported in part by NSF grant AST-2540180. A.A.C.S. is supported by the German \textit{Deut\-sche For\-schungs\-ge\-mein\-schaft, DFG\/} in the form of an Emmy Noether Research Group -- Project-ID 445674056 (SA4064/1-1, PI Sander) and acknowledges financial support by the Federal Ministry for Research, Technology and Space (BMFTR) via the Deutsches Zentrum f\"ur Luft- und Raumfahrt (DLR) grant 50 OR 2509 (PI Sander). A.A.C.S. further acknowledges financial support via the DLR grant 50 OR 2306 (PI Ramachandran/Sander). This project was co-funded by the European Union (Project 101183150 - OCEANS). A.T. acknowledges support from the BELgian federal Science Policy Office (BELSPO) through PRODEX grant PLATO (ZKE8588), from the Flemish Government under the long-term structural Methusalem funding program by means of the project SOUL: Stellar evolution in full glory, grant METH/24/012 at KU Leuven, and from the Research Foundation – Flanders (FWO) (grant agreement G0ABL24N). J.E.M.-D. gratefully acknowledges support from the Secretar\'ia de Ciencia, Humanidades, Tecnolog\'ia e Innovaci\'on (SECIHTI), project CBF-2025-I-2048, ''Resolviendo la F\'isica Interna de las Galaxias: De las Escalas Locales a la Estructura Global con el SDSS-V Local Volume Mapper,'' and from the UNAM/DGAPA/PAPIIT project IA103326, ''DESIRED (DEep Spectra of Ionized REgions Database): de las emisiones m\'as sutiles a la f\'isica fundamental del universo.''

Funding for the Sloan Digital Sky Survey V has been provided by the Alfred P. Sloan Foundation, the Heising-Simons Foundation, the National Science Foundation, and the Participating Institutions. SDSS acknowledges support and resources from the Center for High-Performance Computing at the University of Utah. SDSS telescopes are located at Apache Point Observatory, funded by the Astrophysical Research Consortium and operated by New Mexico State University, and at Las Campanas Observatory, operated by the Carnegie Institution for Science. The SDSS web site is \url{www.sdss.org}.
SDSS is managed by the Astrophysical Research Consortium for the Participating Institutions of the SDSS Collaboration, including the Carnegie Institution for Science, Chilean National Time Allocation Committee (CNTAC) ratified researchers, Caltech, the Gotham Participation Group, Harvard University, Heidelberg University, The Flatiron Institute, The Johns Hopkins University, L'Ecole polytechnique f\'{e}d\'{e}rale de Lausanne (EPFL), Leibniz-Institut f\"{u}r Astrophysik Potsdam (AIP), Max-Planck-Institut f\"{u}r Astronomie (MPIA Heidelberg), Max-Planck-Institut f\"{u}r Extraterrestrische Physik (MPE), Nanjing University, National Astronomical Observatories of China (NAOC), New Mexico State University, The Ohio State University, Pennsylvania State University, Smithsonian Astrophysical Observatory, Space Telescope Science Institute (STScI), the Stellar Astrophysics Participation Group, Universidad Nacional Aut\'{o}noma de M\'{e}xico, University of Arizona, University of Colorado Boulder, University of Illinois at Urbana-Champaign, University of Toronto, University of Utah, University of Virginia, Yale University, and Yunnan University. 

The LBT is an international collaboration among institutions in the United States and Europe. At the time data were acquired for this research, LBT Corporation Members were the University of Arizona on behalf of the Arizona Board of Regents; Istituto Nazionale di Astrofisica, Italy; LBT Beteiligungsgesellschaft, Germany, representing the Max-Planck Society, the Leibniz Institute for Astrophysics Potsdam, and Heidelberg University; and The Ohio State University, representing The Ohio State University, University of Notre Dame, University of Minnesota, and University of Virginia.  This research used the facilities of the Italian Center for Astronomical Archives (IA2) operated by INAF at the Astronomical Observatory of Trieste. Observations have benefited from the use of ALTA Center (alta.arcetri.inaf.it) forecasts performed with the Astro-Meso-Nh model. Initialization data of the ALTA automatic forecast system come from the General Circulation Model (HRES) of the European Centre for Medium Range Weather Forecasts.

\end{acknowledgements}

%
   \bibliographystyle{aa} 
   \bibliography{references.bib} 
%

\appendix

\section{Observation overview}
\label{appendix:obs_overview}

In Table~\ref{tab:obs_overview} we provide an overview of the optical observations, including the measured RVs and the S/N estimates. 
Table~\ref{tab:phot_overview} lists archival flux measurements from \textit{Gaia}, 2MASS, and WISE, which were used in the photometric analysis.

\setlength{\extrarowheight}{3pt}
\begin{table}
\caption{Summary of spectroscopic observations.}\label{tab:obs_overview}
\centering
\begin{tabular}{c c c c c}
\toprule
Instrument & MJD & S/N & RV$_1$ & RV$_2$ \\
 & (d) & &  (km\,s$^{-1}$) & (km\,s$^{-1}$) \\
\midrule
BOSS & 60636.32 & 35 & $-103\pm21$ &  $60\pm29$  \\
BOSS & 60954.50 & 31 & $-96\pm45$ & $-196\pm41$ \\
BOSS & 60986.38 & 31 & $-97\pm19$ & $35\pm25$ \\
\addlinespace
GMOS & 61053.26 & 41 & $-89\pm10$ &  $20\pm21$ \\
GMOS & 61055.21 & 49 & $-82\pm10$ & $-212\pm20$  \\
GMOS & 61059.22 & 51 & $-87\pm11$ &  $22\pm21$  \\
GMOS & 61061.20 & 55 & $-31\pm11$ &  $-208\pm21$ \\
\addlinespace
HIRES & 61002.45 & 10 & $-77\pm19$ & $-230\pm26$ \\
HIRES & 61058.25 & 23 & $-118\pm9$ & $72\pm8$ \\
\addlinespace
PEPSI & 61028.15 & 43 & $-125\pm5$ & $46\pm7$  \\
\bottomrule
\end{tabular}
\tablefoot{Columns list the instrument, observation date (MJD), S/N at 5750\AA, and measured RVs and uncertainties for the primary star (from He~I and H~Balmer absorption) and the companion (from \ion{He}{ii} and \ion{N}{iv} emission).}
\end{table}
\setlength{\extrarowheight}{0pt}

\setlength{\extrarowheight}{3pt}
\begin{table}
\caption{Overview of literature photometric flux measurements used for the SED analysis.}\label{tab:phot_overview}
\centering
\begin{tabular}{c c c c}
\hline
        FilterID & $\lambda_\mathrm{eff}$ & Flux & Flux Error \\
                 & [\AA] &\multicolumn{2}{c}{[erg/cm$^2$/s/\AA]} \\
        \hline
        GAIA3.Gbp & 5035.75 &  3.758E-15 & 0.011E-15 \\
        GAIA3.G & 5822.39 &  6.229E-15 & 0.016E-15\\
        GAIA3.Grp & 7619.96 &  8.166E-15 & 0.030E-15\\
        \addlinespace
        2MASS.J & 12350 & 8.01E-15 & 0.17E-15\\
        2MASS.H & 16620 & 4.276E-15 & 0.091E-15\\
        2MASS.Ks & 21590 & 2.035E-15 & 0.034E-15 \\
        \addlinespace
        WISE.W1 & 33526 & 4.85E-16 & 0.11E-16 \\
        WISE.W2 & 46028 & 1.609E-16 &  0.030E-16\\
        WISE.W3 & 115608 & 6.17E-18 & 0.33E-18\\
        WISE.W4 & 220883 & 1.82E-18 & (\textit{upper limit}) \\
        
        \hline
\end{tabular}
\tablefoot{Columns list filter names, effective wavelengths, fluxes and flux uncertainties.}
\end{table}
\setlength{\extrarowheight}{0pt}

\section{Photometric variability}
\label{appendix:light_curve}

\begin{figure}[t]
    \centering
    \includegraphics[width=\columnwidth]{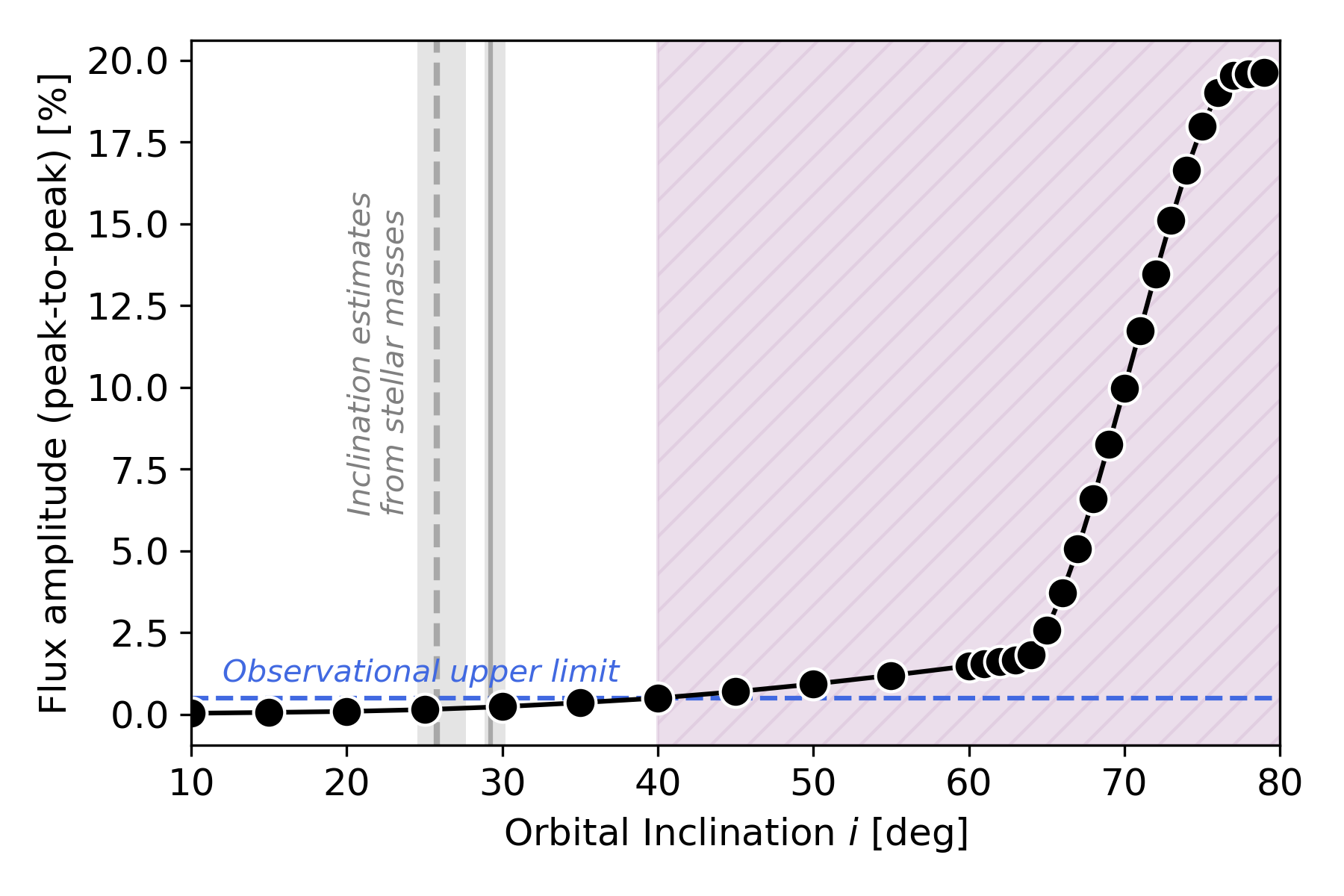}
    \caption{Predicted peak-to-peak photometric variability as a function of orbital inclination, computed with \texttt{PHOEBE} using the stellar and orbital parameters inferred in this work. The horizontal dashed line marks the approximate upper limit of $\sim0.5\%$ variability inferred from the TESS light curve. Inclinations above $\sim40^\circ$ (shaded purple) are expected to produce significantly stronger variability, whereas the inclination estimate of $26^\circ$--$29^\circ$ inferred from the orbit and evolutionary/spectroscopic mass of the primary (solid/dashed grey) is consistent with the observed light curves.}
    \label{fig:lightcurve}
\end{figure}

We further inspected archival photometric data of the target from the Zwicky Transient Facility \citep[ZTF;][]{Bellm+2019, Masci+2019} and the Transiting Exoplanet Survey Satellite (TESS). For the latter, we retrieved the TESS Gaia Light Curve \citep[TGLP;][]{Han+2023} for TIC~53809126, which is available over six observing sectors. Neither the TESS nor the ZTF light curves show obvious periodic variability with amplitudes exceeding approximately $0.5\%$ on timescales shorter than the orbital period of 5.94\,d. In particular, we find no evidence for eclipses or ellipsoidal variability. We note that the TESS light curves were analysed also by \citet{Green+2023} but the star not identified as an ellipsoidal binary candidate.

As a consistency check, we computed synthetic light curves with the \texttt{PHOEBE} code \citep{Prsa+2005,Prsa+2016,Horvat+2018,Jones+2020,Conroy+2020}. We adopted blackbody atmospheres for both stars, computed the light curves in the TESS passband, and fixed the orbital period to the measured value. The stellar temperatures and radii were taken from the spectral analysis, while the component masses were varied consistently with the inclination-dependent dynamical solution.
Figure~\ref{fig:lightcurve} shows the predicted peak-to-peak photometric variability as a function of orbital inclination. At high inclinations, eclipses produce increasingly large photometric amplitudes. The absence of variability above approximately $0.5\%$ is consistent with an inclination of $i \lesssim 40^\circ$. This agrees with the inclination estimate of $26^\circ$--$29^\circ$ from the dynamical masses (Sect.~\ref{sec:stellar_masses}).

\section{Radial velocities}
\label{appendix:todcor}

\begin{figure}[t]
    \centering
    \includegraphics[width=\columnwidth]{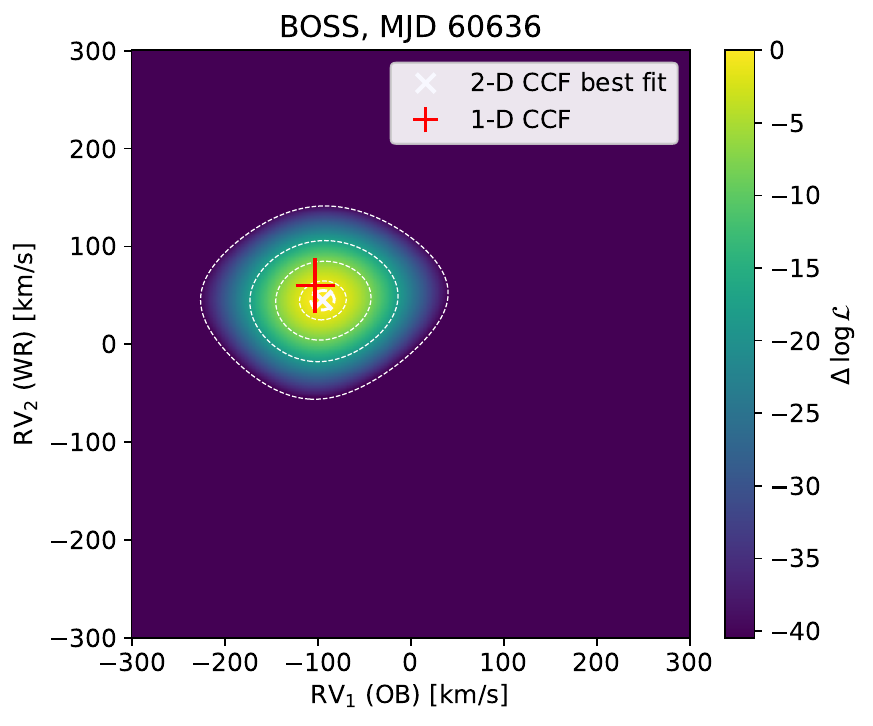}
    \caption{Component RVs for the BOSS spectrum (MJD 60636) inferred from two-dimensional template cross-correlation. The figure shows the likelihood surface for the RVs of the primary O-type star vs. the companion star with WR-like spectrum. The best-fit values are marked with a white cross, a red marker indicates the values inferred from one-dimensional cross-correlation.}
    \label{fig:todcor_example}
\end{figure}

The RVs used throughout this work were derived from one-dimensional template cross-correlation (Sect.~\ref{sec:radial_velocities}), where wavelength regions dominated by one of the two stellar components were analysed independently. To verify that this approach does not introduce significant biases, we repeated the RV determination using the final best-fit model spectra derived from the spectral analysis (Sect.~\ref{sec:spectral_fit_results}) and performed a simultaneous two-dimensional cross-correlation.
We adopted the \texttt{TODCOR} algorithm by \citet{Mazeh_Zucker1994} and \citet{Zucker2021}, recast as a Gaussian-likelihood fit. The likelihood was evaluated on a grid of primary and secondary RVs, and the maximum-likelihood solution was adopted as the best-fitting pair of velocities. Figure~\ref{fig:todcor_example} shows an example of the resulting likelihood surface for the BOSS spectrum obtained on MJD 60636.
The RVs obtained with \texttt{TODCOR} agree with those derived from the one-dimensional cross-correlation. For all spectra, except for the PEPSI observation, the differences are smaller than $2\sigma$. The PEPSI spectrum shows larger formal discrepancies, while the absolute differences remain modest (14 and $26\,\mathrm{km\,s^{-1}}$ for the primary and secondary, respectively). Overall, the consistent results indicate that the simpler one-dimensional analysis provides reliable RV measurements for this system.

\section{Hydrogen fraction}
\label{appendix:hydrogen_frac}

\begin{figure}[t]
    \centering
    \includegraphics[width=\columnwidth]{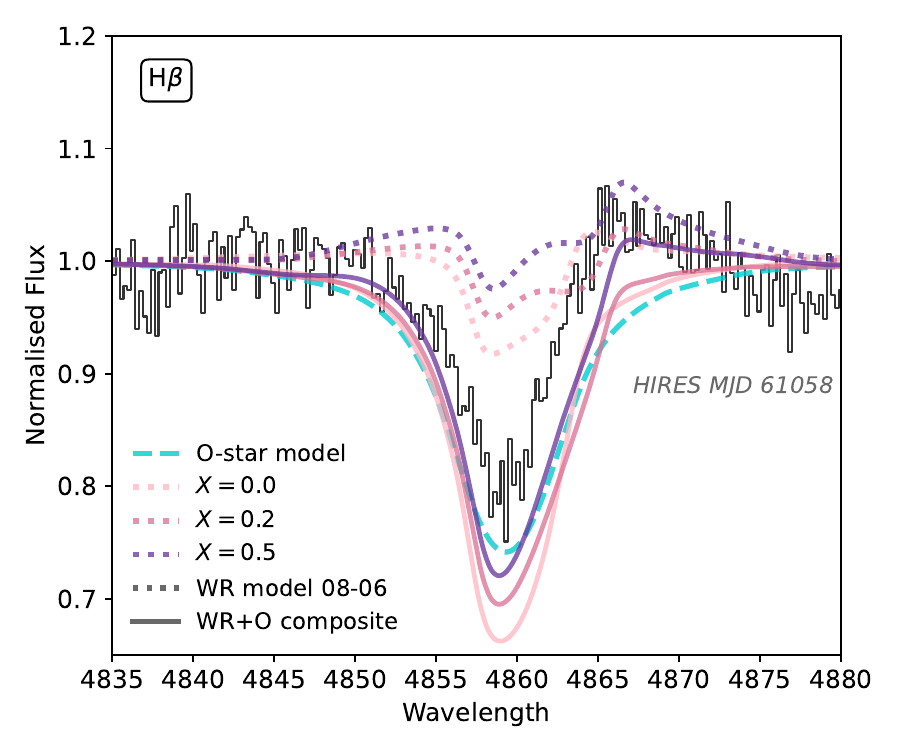}
    \caption{Comparison of PoWR model spectra with different hydrogen mass fractions around the \ion{H}{$\beta$} line. The observed composite HIRES spectrum (MJD 61058) is shown in black, while coloured curves show models with identical stellar temperature and transformed radius but hydrogen mass fractions of $X=0.0$, $0.2$, and $0.5$. The combined model spectra (solid lines) are composites of the O-star model (cyan dashed) and the WR model spectra (dotted). Increasing the hydrogen abundance strengthens the Balmer emission. The model with $X=0.5$ provides the best qualitative agreement with the observations and was adopted for the tailored model grid.}
    \label{fig:hydrogen_frac_comparison}
\end{figure}

The hydrogen mass fraction of the stripped star was not included as a free parameter in the tailored PoWR model grid, but was fixed at $X=0.5$. This choice was guided by comparisons with the publicly available PoWR Wolf--Rayet grids \citep[$X=0.0$, 0.2, and 0.5;][]{Hamann+2004,Todt+2015}, which span a range of surface hydrogen abundances.
Figure~\ref{fig:hydrogen_frac_comparison} illustrates the effect of varying the hydrogen abundance on the \ion{H}{$\beta$} line for models with fixed stellar temperature and transformed radius. Models with little or no hydrogen substantially underpredict the observed Balmer emission, whereas the model with $X=0.5$ provides the best qualitative agreement with the spectrum. We find the same overall behaviour when varying the stellar temperature and transformed radius within the range favoured by the spectral analysis. These comparisons therefore indicate that the stripped star retains a substantial amount of hydrogen at its surface, motivating our adoption of $X=0.5$ for the tailored PoWR$^{\textsc{hd}}$ model grid.
We defer a detailed abundance analysis to future work.

\section{Literature comparison}
\label{appendix:lit_comparison}

\begin{figure*}[t]
    \centering
    \includegraphics[width=\textwidth]{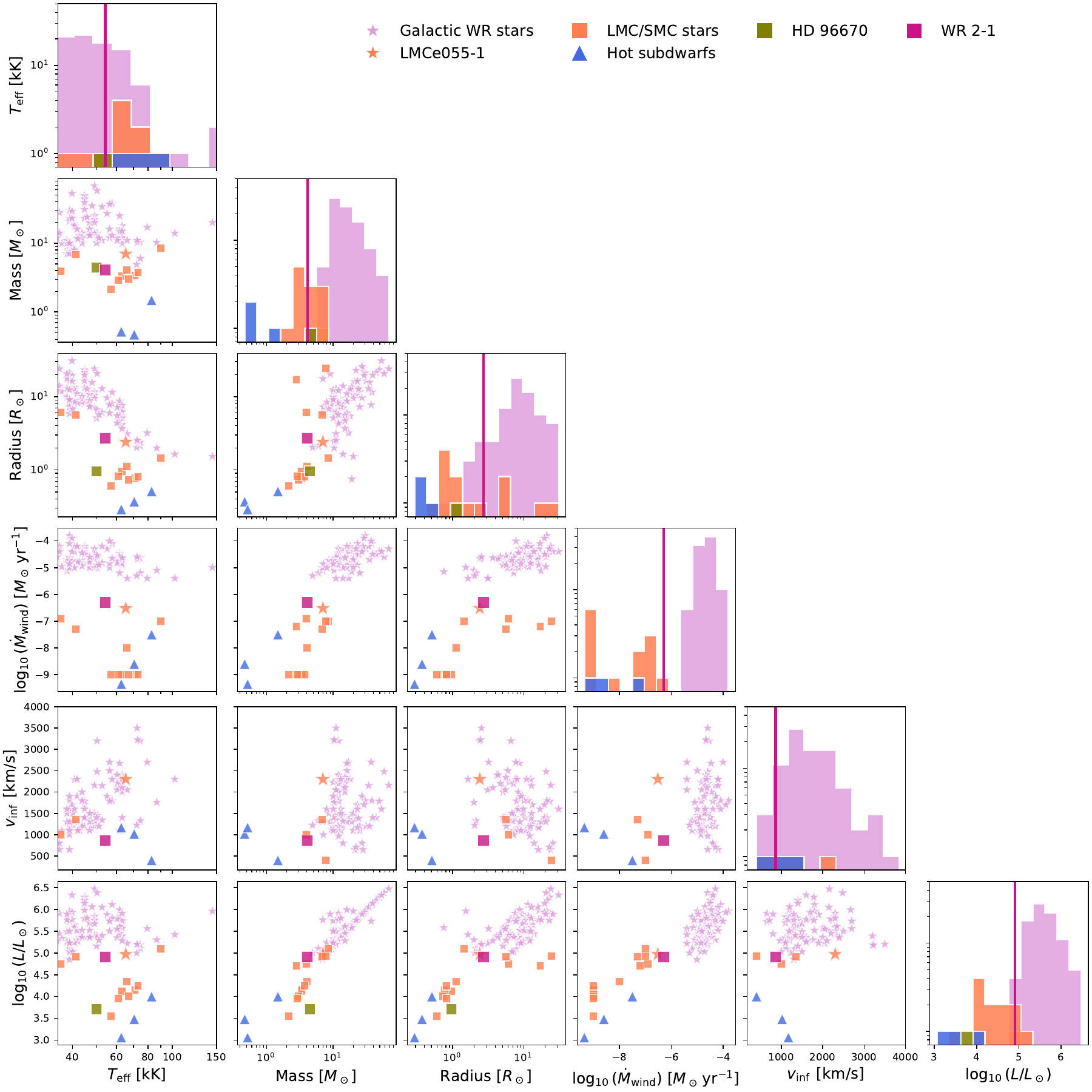}
    \caption{Corner plot comparing stellar parameters of WR\,2-1 to known stripped stars from literature. The figure includes Galactic WR stars \citep[light pink;][]{Sander+2012,Hamann+2019,Sander+2019, Hamann+2006, Todt+2010,Zhang+2020,Massey+2024,Lefever+2026}, high-temperature hot subdwarfs and merger candidates \citep[blue;][]{Krticka+2024,Mueller-Horn+2026}, intermediate-mass stripped stars in the Magellanic Clouds \citep[orange;][]{Drout+2023,Goetberg+2023,Ramachandran+2023,Ramachandran+2024, Massey+2024}, and the candidate intermediate-mass stripped star HD~96670 \citep[olive;][]{Naze_Rauw2025}. The panels show effective temperature, stellar mass, stellar radius, logarithmic mass loss rate, terminal wind speed and logarithmic luminosity (left to right). In most aspects, WR\,2-1 is intermediate between classical WR stars and hot subdwarfs.}
    \label{fig:lit_comparison_corner}
\end{figure*}

In Fig.~\ref{fig:lit_comparison_corner}, we show a multi-parameter comparison between WR\,2-1 and known stripped stars. The corner plot features two-dimensional scatter plots and marginalised distributions of stellar temperatures, masses, radii, mass loss rates, terminal wind velocities, and luminosities, where available. WR\,2-1 generally occupies the parameter space intermediate between hot subdwarf stars and the WR regime. It most closely resembles the intermediate-mass stripped stars in the Magellanic Clouds, and in particular the low-luminosity LMC WR star LMCe055-1. 

\section{Evolutionary models}
\label{appendix:mesa_set-up}

\begin{figure*}[t]
    \centering
    \includegraphics[width=0.95\textwidth]{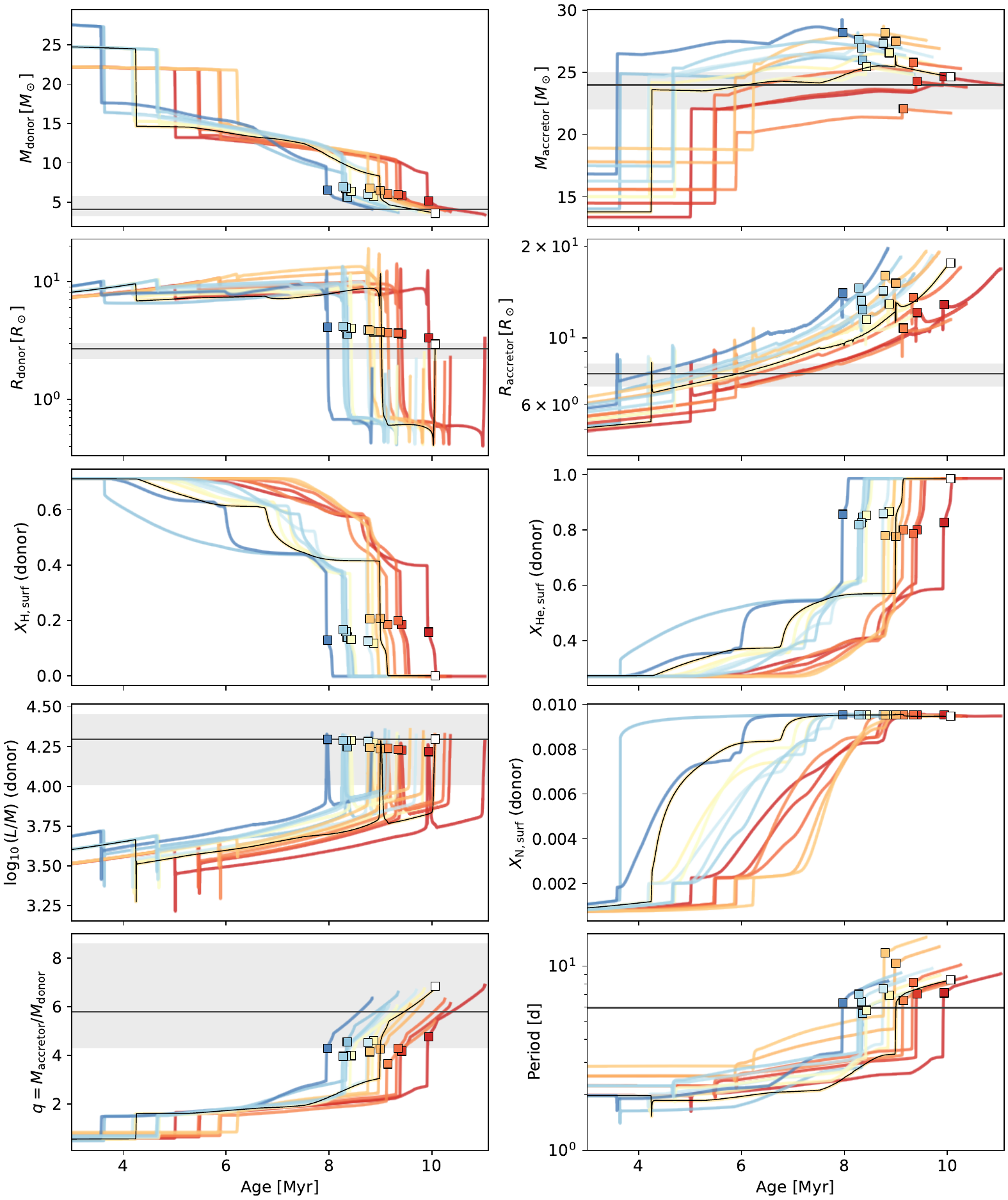}
    \caption{Time evolution of stellar parameters for simulated binary models compared with observed properties of WR\,2-1. The evolutionary tracks are shown for selected models from the grid of \citet{Jin+2026} (coloured lines, see Sect.~\ref{sec:evolution}). The representative model from Fig.~\ref{fig:mesa_model_HRD} is highlighted in black. The epoch that best matches the observed properties in terms of donor temperature, luminosity, and mass and binary mass ratio is indicated by square markers for all models. Horizontal black lines and shaded regions denote the measured parameters of WR\,2-1, where available. The different panels show stellar masses, radii, surface hydrogen, helium, and nitrogen mass fractions of the donor star, luminosity-to-mass ratio of the donor, the binary mass ratio, and the period evolution.}
    \label{fig:mesa_model_time}
\end{figure*}

Figure~\ref{fig:mesa_model_time} compares the observed properties of WR\,2-1 with selected binary evolution models from the grid of \citet{Jin+2026} (see Sect.~\ref{sec:evolution}). The models were chosen because they approximately reproduce the present-day properties of WR\,2-1 at some stage of their evolution. The panels show the evolutionary tracks of the stellar masses, radii, surface hydrogen, helium, and nitrogen mass fractions of the donor star, the donor luminosity-to-mass ratio, the mass ratio, and the orbital period. The representative model shown also in Fig.~\ref{fig:mesa_model_HRD} is highlighted in black. Square symbols indicate the evolutionary stage that provides the closest match to the observed system parameters; these solutions are distributed roughly equally between models before and after core-helium burning. Horizontal black lines and shaded regions denote the measured properties of WR\,2-1, where available.
Overall, the models reproduce most of the observed system properties reasonably well. However, at the evolutionary stage that best matches the stripped-star component, the accretor is generally predicted to be somewhat more evolved than observed, with a higher luminosity and larger radius.

\end{document}